\documentclass[10pt,journal,compsoc]{IEEEtran}

\usepackage{booktabs} 
\usepackage{graphicx}
\usepackage[singlespacing]{setspace}
\usepackage{threeparttable}
\usepackage{multirow}
\usepackage{subcaption}
\usepackage[utf8]{inputenc}
\usepackage[english]{babel}
\usepackage{amsthm}
\usepackage{amsmath}
\usepackage{verbatim}
\DeclareFontFamily{OT1}{pzc}{}
\DeclareFontShape{OT1}{pzc}{m}{it}{<-> s * [1.1] pzcmi7t}{}
\DeclareMathAlphabet{\mathpzc}{OT1}{pzc}{m}{it}
\theoremstyle{definition}
\newtheorem{definition}{Definition}
\DeclareMathAlphabet\mathbfcal{OMS}{cmsy}{b}{n}
\usepackage{centernot}
\usepackage{enumitem}
\usepackage{stfloats}
\RequirePackage[prologue,dvipsnames]{xcolor}

\usepackage[hyphens]{url}
\usepackage[hidelinks]{hyperref}
\hypersetup{breaklinks=true}

\usepackage[ruled, vlined, linesnumbered]{algorithm2e}

\SetCommentSty{mycommfont}

\usepackage{amsmath, listings, amsthm, amssymb, proof, xspace}
\usepackage{tikz}
\usetikzlibrary{automata,positioning}
\usetikzlibrary{shapes.geometric, arrows}
\usetikzlibrary{fit,arrows,calc}
\usetikzlibrary{shadows}
\usetikzlibrary{external}
\newcommand\token[2] {
\tikzexternaldisable
\hspace*{-2pt}\tikz{\node[circle, drop shadow, fill=#2, text=white, inner sep=1pt, font=\normalsize, align=center, midway] {#1};}\hspace*{-1pt}
\tikzexternalenable
}
\lstdefinelanguage{execrule}{
  keywords={rule, select, where, send, reply, when},
  keywordstyle=\color{Purple}\ttfamily\bfseries,
  keywords=[2]{applies, when, random, state, transition, component,dbg,receipt,:,|,;},
  keywordstyle=[2]\color{black}\ttfamily\bfseries,
  sensitive=false,
  comment=[l]{//},
  morecomment=[s]{/*}{*/},
  commentstyle=\color{green}\ttfamily,
  stringstyle=\color{red}\ttfamily,
  morestring=[b]',
  morestring=[b]"
}

\lstdefinelanguage{antlr}{
  keywords={grammar, rule,ExecRule,EOF},
  keywordstyle=\color{Purple}\ttfamily\bfseries\footnotesize,
  keywords=[2]{script:, execRule:,when:, where:, statement:, scriptStatement:,interactiveStatement:,umlrtCmd:,random:,dbgCmd:,|},
  keywordstyle=[2]\color{black}\ttfamily\bfseries\footnotesize,
  otherkeywords = {-,|},
  morekeywords = [3]{-},
  morekeywords = [4]{|},
  keywordstyle = [3]{\color{blue}}\footnotesize,
  keywordstyle = [4]{\color{blue}}\footnotesize,
  identifierstyle=\color{black}\footnotesize,
  sensitive=true,
  comment=[l]{//},
  morecomment=[s]{/*}{*/},
  commentstyle=\color{green}\ttfamily,
  stringstyle=\color{blue}\ttfamily,
  morestring=[b]',
  morestring=[b]",
  numbers=left,
  xleftmargin=2em,
  frame=single,
  framexleftmargin=1.5em,
  numberstyle=\color{black}\footnotesize\ttfamily,
  alsodigit={:}
}

\theoremstyle{remark}

\ifCLASSOPTIONcompsoc
  \usepackage[nocompress]{cite}
\else
  \usepackage{cite}
\fi

\ifCLASSINFOpdf
\else
\fi

\usepackage{graphicx}
\usepackage[singlespacing]{setspace}
\usepackage{threeparttable}
\usepackage{multirow}
\usepackage[splitrule]{footmisc}

\begin{document}

\title{Execution of Partial State Machine Models}
%
%
%
%

\author{Mojtaba~Bagherzadeh,\thanks{Mojtaba Bagherzadeh and Nafiseh Kahani are  with the school of EECS,
University of Ottawa.}\thanks{Karim Jahed and Juergen Dingel are with the School of Computing, Queen’s University, Canada.}\thanks{E-mail: m.bagherzadeh@uottawa.ca, nkahani@uottawa.ca, jahed@cs.queensu.ca, dingel@cs.queensu.ca} Nafiseh Kahani, Karim Jahed, Juergen Dingel}
\markboth{}%
{Shell \MakeLowercase{\textit{et al.}}: Bare Advanced Demo of IEEEtran.cls for IEEE Computer Society Journals}

\IEEEtitleabstractindextext{%
\begin{abstract}
The iterative and incremental nature of software development using models typically makes a model of a system incomplete (i.e., partial) until a more advanced and complete stage of development is reached. Existing model execution approaches (interpretation of models or code generation) do not support the execution of partial models. Supporting the execution of partial models at early stages of software development allows early detection of defects, which can be fixed more easily and at lower cost. 

This paper proposes a conceptual framework for the execution of partial models, which consists of three steps: \textit{static analysis}, \textit{automatic refinement}, and \textit{input-driven execution}. First, a static analysis that respects the execution semantics of models is applied to detect problematic elements of models that cause problems for the execution. Second, using model transformation techniques, the models are refined automatically, mainly by adding decision points where missing information can be supplied. Third, refined models are executed, and when the execution reaches the decision points, it uses inputs obtained either interactively or by a script that captures how to deal with partial elements.

We created an execution engine called \textit{PMExec} for the execution of partial models of UML-RT (i.e., a modeling language for the development of soft real-time systems) that embodies our proposed framework. We evaluated \textit{PMExec} based on several use-cases that show that the static analysis, refinement, and application of user input can be carried out with reasonable performance, and that the overhead of approach, which is mostly due to the refinement and the increase in model complexity it causes, is manageable. We also discuss the properties of the refinement formally, and show how the refinement preserves the original behaviors of the model.

\end{abstract}

}

\maketitle

\IEEEdisplaynontitleabstractindextext
%
\IEEEpeerreviewmaketitle

\section{Introduction}
\label{sec:introduction}

\IEEEPARstart{M}odel Driven Engineering (MDE) is a software development methodology that advocates the use of model for the description and development of systems~\cite{MDE1}. These models can capture relevant concepts on a level of abstraction higher than source code, and thus can facilitate communication, automation, and reuse. In general, the level and form of the use of models varies greatly, from the use of models only for documentation and communication to the use of models for code generation and as the main software development artifacts rather than source code~\cite{whittle2013state,mohagheghi2008proof,mohagheghi2013empirical,torchiano2013relevance}. While modeling is used in a range of industries such as telecom, automotive, aerospace, business, and  military, its use for the development of Real-time Embedded (RTE) systems appears to be one of the most prevalent~\cite{MDEINRTE1,MDEINRTE2,MDEINRTE3}. 

Over the last two decades, an impressive number of MDE tools and techniques for RTE systems have been introduced such as IBM RSARTE~\cite{RSART}, ANSYS SCADE Suite~\cite{SCADE}, YAKINDU Statecharts~\cite{YAKINDU}, and AUTOFocus~\cite{AutoFocus}. These tools provide a range of capabilities to simplify the development of RTE systems using models. E.g., the models can be executed, debugged, analyzed, visualized, and transformed. Our work concerns the \emph{execution} of models, which is typically supported either by interpreting the models or by translating them into an existing programming language, often by code generation (translational execution)~\cite{mojtabadebugging,Ciccozzi2018}.

The ability to execute models in some ways is an important capability of MDE tools because it enables many important quality assurance activities such as testing and debugging.
However, while existing MDE tools offer good support for the execution of complete models, none of them make much effort to extend that support to models that are incomplete.
For instance, a state machine model may be incomplete because the behaviour of a component or a composite state has not yet been specified, or the specification of a transition is missing or incomplete (due to, e.g., missing triggers or guards, or incomplete action code).
Although these models might contain many executable parts and there might be great value in the ability to execute them, existing tools typically do not allow this. One reason is that code generation or build operations might fail.
But, even if these steps succeed, the tools do not allow a `best effort' treatment of partial models in which execution proceeds as far as possible and when it cannot proceed any further due missing information, then this missing information can be supplied manually or automatically to allow the execution to continue.
Our work aims to enable this kind of best effort treatment. 

Generally, the benefits of partial model execution can be classified as follows:

\begin{enumerate}
\item {\em Facilitate design decisions.}
In very early stages of development, the ability to execute a partial model may help perform design space exploration, evaluate design alternatives and explore tradeoffs in a more efficient, focused fashion and without having to flesh out details that are irrelevant to the design decision, but required to achieve executability.
The goal is to allow the model to be useful as early as possible and with a minimum of procedural or notational accidental complexity~\cite{selic2012will}.

\item {\em Facilitate validation and improvement of models.}
A basic tenet in software engineering is that development should facilitate the early detection of bugs, because the cost of fixing a bug tends to increase with the amount of time that it goes unnoticed~\cite{earlydefect1,earlydefect2,earlydefect3}.
Agile development activities such as continuous testing are motivated by this observation. 
In the context of MDE, this means that developers should be able to carry out validation and debugging activities as early as possible and not only after additional effort has been invested to make the model complete.
The ability to execute partial models is necessary to achieve this vision and a key prerequisite for making MDE more agile~\cite{selic2012will}.

But, partial model execution can also help with the validation of large, complete models. In large MDE projects, the system can have hundreds of components~\cite{heijstek2009empirical,UMLRTUse}, much reducing the feasibility and practicality of system-wide validation activities, especially when execution requires code generation (e.g., as reported in~\cite{inccodegeneration} the code generation of large systems can take hours to complete).
In these settings, unit testing of individual components is required.
Each such component $c$ is a partial model that typically is not executable in isolation.
To achieve executability, the current state-of-the-art demands that $c$ be completed by a harness that mocks or stubs out the parts of the system that $c$ relies on.
The creation of a suitable harness and connecting it to $c$ can involve a significant amount of accidental complexity.
Our approach facilitates unit testing, because the developer can focus on supplying appropriate information to $c$ if and when $c$ needs it. 

\item {\em Facilitate collaborative, heterogeneous development}.
MDE is often collaborative and different model parts can be owned by different, possibly geographically distributed, teams~\cite{MDEINRTE1Journal}.
As a result, components may be out-of-date, incorrect, or unavailable which may affect the executability of any other components using them.
Partial model execution can help protect the developers of a part of the system from these issues by allowing them to, e.g., perform validation without having to wait for a new version or attempt to make their part work with an old, out-dated version. 

But, MDE can also be part of a larger, heterogeneous development process in which, e.g., the code generated from a model is integrated with other code that has been developed by a different team using a different process or tool, or been purchased from a vendor~\cite{hutchinson2011empirical,MDEINRTE1Journal}.
Partial model execution can help here, too, because the model can be validated without having to obtain this additional code and integrate it with the model.

According to surveys, the integration of MDE into industrial development processes can be challenging, especially when distributed development and interoperability with existing code or tools is required~\cite{mohagheghi2008proof,hutchinson2011empirical,MDETooling1,MDEINRTE1Journal}.
Partial model execution can increase the flexibility of MDE and thus may help mitigate these problems and improve industrial adoption of MDE. 
\end{enumerate}

Despite the importance and benefits of the execution of partial models, no work has addressed the execution of partial models, to the best of our knowledge. Existing work in this context deals with partial models at design time, which allows for specification, analysis, verification, and transformation of partial models (e.g.,~\cite{partialmodelverify,Famelis:partial,partialmodelspecification,partialmodeltransform}). As mentioned, a possible approach to executing partial models is the simulation of missing components by techniques such as mocking or stubbing~\cite{mockingref1}. These solutions are mainly designed for unit testing, and have several deficiencies when used for the debugging of models: (1) They are not fully automated, and developers still need to do extra work to, e.g., create stubs for the missing components. (2) Often, they are applicable only at the component-level to simulate a component fully, while for debugging purposes, developers may need to simulate only parts of a component. (3) More importantly, while in the code-base development context, there are several mocking frameworks (e.g., Mockito, EasyMock, JMock, Opmock, etc.) that can be used to simulate components of a system~\cite{mocking2}, there is a lack of facilities, guidelines, and frameworks in the context of MDE to help to create mockers~\cite{mockingsemantic}. 

This work advances the state of the art in model-level execution and utilizing partial models by providing support for the execution of partial models. We propose a conceptual framework for the execution of partial models, which consists of three steps:  \textit{static analysis}, \textit{automatic refinement}, and \textit{input-driven execution}. First, a static analysis that respects the execution semantics of models is applied. It detects problematic elements that prevent the execution from progressing or reaching certain states. Second, to make the partial models executable, model-to-model (M2M) transformations~\cite{Kahani2018} are used to refine the models automatically by adding decision points where the elements are missing. The refined models preserve the original behavior of the user-defined models and its execution does not not get stuck and can reach all defined states in a finite number of execution steps, assuming proper inputs are provided. Third, during the execution, these decision points allow users to interactively (1) inspect and modify the system using debugging services, and (2) select one of the possible options to continue the execution. The interactive execution requires manual intervention, which can be repetitive, tedious, and time-consuming. To mitigate this problem, our approach includes a scripting language that captures user input as execution rules that can be applied automatically during execution, without stopping the execution and interacting with users. {Also, the approach allows user input to be saved to the script of the execution rules or the design model, to avoid any unnecessary duplication by minimizing the effort required for writing the script and completion of the design model.}

We extended our previous work on model-level debugging \cite{MDebuggerDemo,mojtabadebugging} in the context of UML-RT (i.e., a  language for modeling of soft real-time systems)~\cite{selic1994real}, and created an execution engine of partial UML-RT models (\textit{PMExec}~\cite{ASE-Paper-demo}) that embodies the proposed framework. To maximize the impact of our work, our implementation is publicly available, and only uses \textit{open source} tools, including the Papyrus-RT MDE tool, for modelling and code generation, and the Epsilon~\cite{epsilon2008} tools for model transformation. We evaluated \textit{PMExec} based on several use-cases that show that the static analysis, refinement, and handling of user input were performed with reasonable quality, and the overhead of approach, which is caused by the increase of complexity of models by the refinement, is manageable.

The rest of this paper is organized as follows. In Section~\ref{sec:background},
we describe our formalization and a running example. Section~\ref{sec:framework} presents
a conceptual framework for execution of partial models, and  Section~\ref{sec:application} discusses the application of the framework into UML-RT. We present our evaluation approach and results in Section~\ref{sec:discussion} and discuss the limitations and issues of our solution in Section~\ref{sec:evaluation}. We review related work in Section~\ref{related-work}, and then conclude the paper with a discussion, summary, and directions for future research.

\newcommand{\codee}[1]{\small\textit{#1}\normalsize}

\newcommand{\code}[1]{\small\texttt{#1}\normalsize}

\section{Preliminaries} \label{sec:background}
In this section, we define, exemplify, and discuss the formalization that we will use to specify and justify our refinement approach. To be able to illustrate our approach, we use the UML profile for Real-time systems (UML-RT).
UML-RT~\cite{selic1994real,Ernesto} is a language specifically designed for Real-Time Embedded (RTE) systems, with soft real-time constraints. Over the past two decades, it has been used successfully in industry to develop several large-scale industrial projects (e.g.,~\cite{UMLRTUse}), and has a long, successful track record of application and tool support, via, e.g., IBM RSA-RTE~\cite{RSART}, RTist~\cite{rtist}, Eclipse eTrice \cite{eTrice} , and Papyrus-RT~\cite{PapyrusRT}. Our formalization is simplified, and focused on aspects that matter most to the execution of partial models. Interested readers can refer to~\cite{Ernesto,selic1994real} for more in-depth information regarding UML-RT.



\newcommand{\alt}{~~|~~}
\newcommand{\comp}[1]{\llbracket #1 \rrbracket}

\newcommand{\inlinexp}[1]{
{\footnotesize
 \[\begin{array}{l}
 #1
 \end{array}\]}}

\newcommand{\inlinexpa}[2]{
{\footnotesize
 \[\begin{array}{#1}
 #2
 \end{array}\]}}

\newcommand{\infr} [3] [] {\infer[\textsc{#1}]{#3}{#2}}
\newcommand{\iand}        {\qquad}

\newcommand{\Ctxt}       {\mathcal{E}}
\newcommand{\InCtxt} [1] {\Ctxt[#1]}


\newcommand{\subst} [3]    {#3 [#2 / #1]}
\newcommand{\dstep} [2]    {#1 ~\Downarrow~ #2}

\newcommand{\ssosredex}        {\rightarrow}
\newcommand{\ctxtreduce}       {\mapsto}
\newcommand{\sstep}     [3] [] {#2 &\ssosredex&  #3 &\textsc{#1}}
\newcommand{\ctxtstep}  [3] [] {#2 &\ctxtreduce& #3 &\textsc{#1}}


\newcommand{\funct} [2] {#1\nobreak\rightarrow\nobreak#2}
\newcommand{\boolt}     {\mathtt{bool}}

\newcommand{\typeEnv}         {\Gamma}
\newcommand{\entails}         {\vdash}
\newcommand{\judgment}   [3] {#1 \entails #2 : #3}
\newcommand{\envent}      [2] {\judgment{\typeEnv}{#1}{#2}}
\newcommand{\extenvent}   [4] {\judgment{\typeEnv, #1 : #2}{#3}{#4}}
\newcommand{\envlookup}   [3] {\infr{#1(#2) = #3}{\judgment{#1}{#2}{#3}}}


\newcommand{\lamdefe}  [2] {\lambda #1.~#2}
\newcommand{\lamdefea} [2] {\begin{array}{l}\lambda#1.\\\hspace*{.5em}#2\\\end{array}}

\newcommand{\letdefe}    [3] {\letbind{#1}{#2}~\letin{#3}}
\newcommand{\letdefarre} [3] {\begin{array}{l}\letbind{#1}{#2}\\\letin{#3})\\\end{array}}

\newcommand{\letbind}  [2] {\mathsf{let}~\lbind{#1}{#2}}
\newcommand{\letbindp} [2] {\mathsf{let}~(\lbind{#1}{#2})}
\newcommand{\lbind}    [2] {#1=#2}
\newcommand{\letin}    [1] {\mathsf{in}~#1}

\newcommand{\ife}      [3] {\ifline{#1}~\thenline{#2}~\elseline{#3}}
\newcommand{\ifea}     [3] {\begin{array}{l}\ifline{#1}\\\thenline{#2}\\\elseline{#3}\end{array}}

\newcommand{\ifop}         {\mathsf{if}}
\newcommand{\ifline}   [1] {\ifop~ #1}
\newcommand{\thenline} [1] {\mathsf{then}~#1}
\newcommand{\elseline} [1] {\mathsf{else}~#1}

\newcommand{\binopdef}     {\mathit{binop}}
\newcommand{\unopdef}      {\mathit{unop}}
\newcommand{\binope}   [2] {\binopdef~#1~#2}
\newcommand{\unope}    [1] {\unopdef~#1}

\newcommand{\andop}        {\mathsf{and}}
\newcommand{\orop}         {\mathsf{or}}
\newcommand{\notop}        {\mathsf{not}}
\newcommand{\ande}     [2] {\mathsf{and}~#1~#2}
\newcommand{\ore}      [2] {\mathsf{or}~#1~#2}
\newcommand{\note}     [1] {\mathsf{not}~#1}

\newcommand{\falsev}     {\mathsf{false}}
\newcommand{\truev}      {\mathsf{true}}



\subsection{A Running Example}
\begin{figure}[t!]
\centering
\includegraphics[width=8cm]{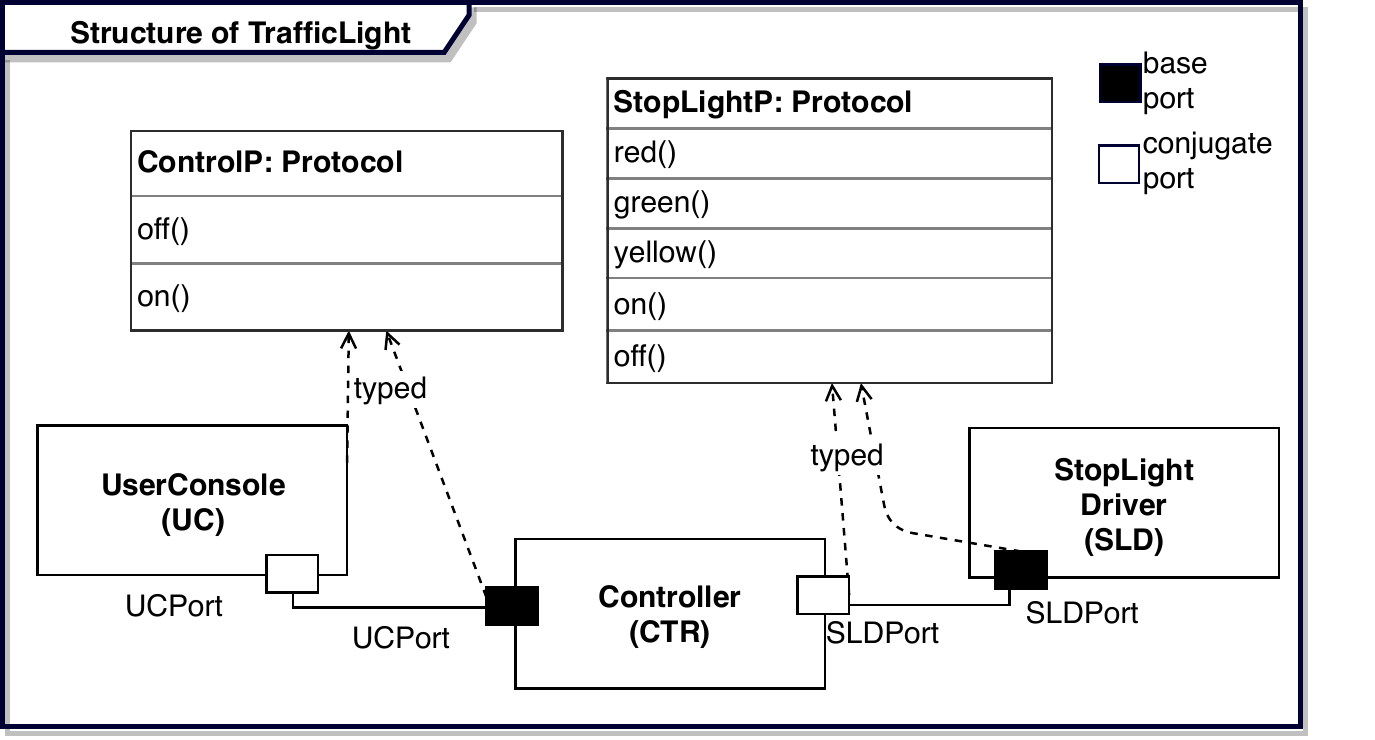}
\caption{The structure of \textit{TrafficLight}}
\label{fig:abmstructure} 
\end{figure}
\begin{figure}[t!]
\centering
\includegraphics[width=8cm]{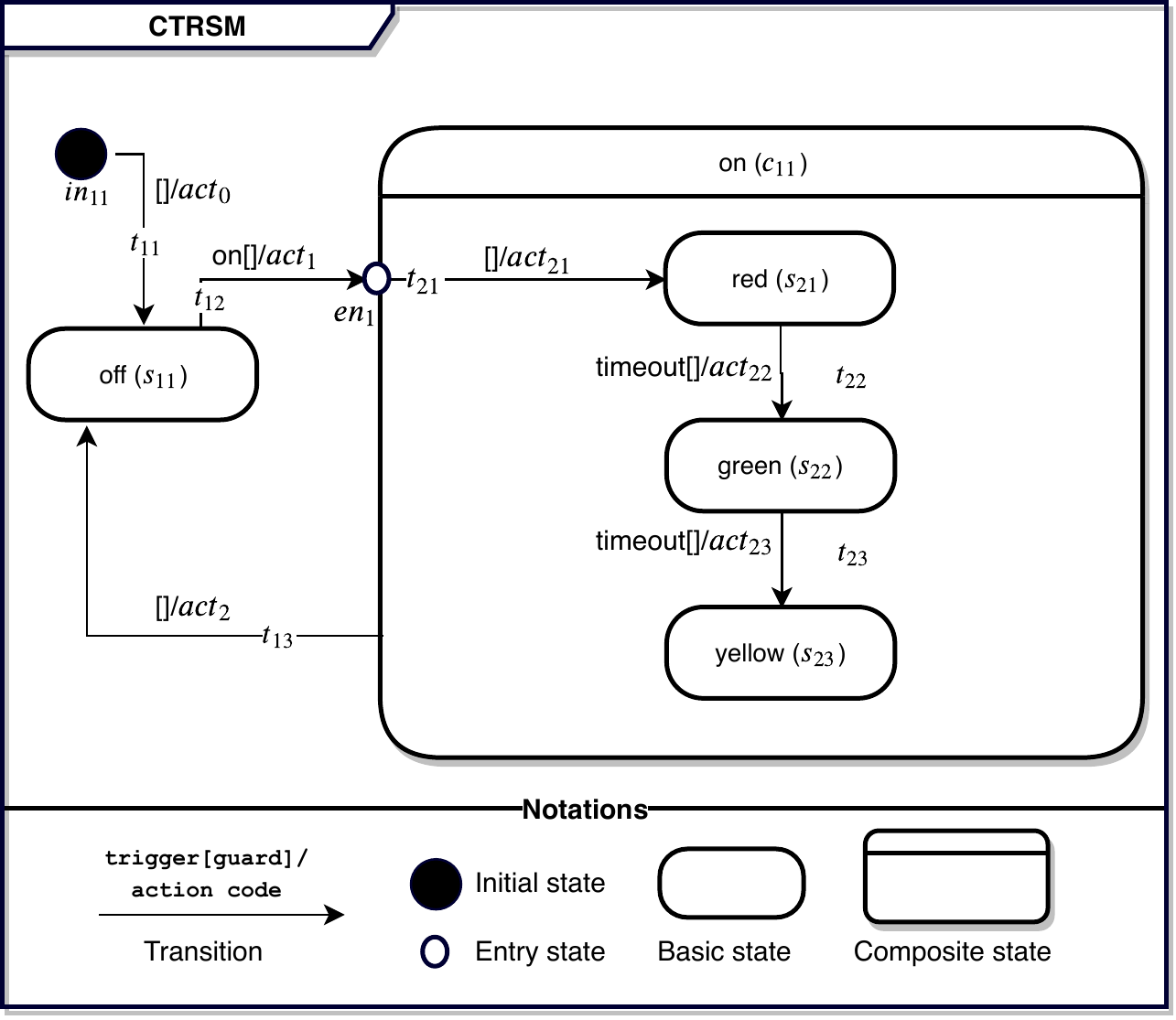}
\caption{Partial Bahaviour of component \textit{CTR} \textit{(CTRHSM)}}
\label{fig:CTRHSM}
\end{figure}
\label{sec:runningexample}
We use the control system of a simple traffic light (\textit{TrafficLight}) as a running example throughout the paper. The structure of the system is shown in Figure~\ref{fig:abmstructure}, which consists of three components: UserConsole \textit{(UC)}, Controller \textit{(CTR)}, and StopLightDriver \textit{(SLD)}.  The \textit{UC} component collects user input, which it passes on to the \textit{CTR} component, the component controlling the light. Using the corresponding messages, the \textit{CTR} component sends the control actions to the \textit{SLD} component, which transfers the messages through a hardware port to the traffic light.

Let us assume that we have a partial model of \textit{TrafficLight} in which the behaviour of the component \textit{UC} is missing, the behaviour of capsule \textit{SLD} is complete, and the behaviour of capsule \textit{CTR} is incomplete as shown in Figure~\ref{fig:CTRHSM} (e.g., no outgoing transition is defined from \textit{yellow}).  Let us discuss two exemplary situations where the execution of partial models can be helpful for early evaluation of design decisions and unit testing. 

\textbf{Scenario-1: (Early evaluation of design decisions)} We want to execute the model as is and test the current design of the component \textit{CTR}. Existing MDE tools do not support the execution of the example out-of-the-box, due to two issues: (1) An appropriate stub for capsule \textit{UC} needs to be created, which is a time-consuming task, (2) even if the stub is  provided, the execution of \textit{CTR} stops at state \textit{yellow}, due to the missing specification, i.e., no outgoing transition is defined from state \textit{yellow}. Currently, the only way to resolve the second issue is to postpone the evaluation until the missing specifications are provided. 

\textbf{Scenario-2: (Unit testing)} We want to perform unit testing of the component \codee{CTR}. To do that, all relevant components except \codee{CTR} should be replaced by appropriate stubs. Note that even though the design of \codee{SLD} is complete, it is not reasonable to use it for unit testing since it requires interaction with hardware which can be time-consuming and complicates the testing process.  

In this work, we will present a systematic solution for automatically creating an executable version of a partial model that allows testing and debugging. For example, using our solution both of the above scenarios can be addressed without a need for the creation of stubs and completion of models.

\subsection{Modelling of a Real-time Embedded (RTE) System Using UML-RT}

\begin{definition}{\textit{(Read function (Projection))}}  \label{def0}
Let $tp$ be a tuple of attributes $\langle a_1, \dots ,a_n \rangle$ where $a_1 \dots a_n$ refer to attributes names.  We use $tp.a_i$  to read the value of attribute $a_i$. E.g., to read the value of attribute $name$ of tuple $person = \langle name, family \rangle$ we can use $person.name$.
\end{definition}

 \begin{definition} \textit{(Interface)}  \label{def:interface}
 Let us define an interface as a set of pairs $({m},{d})$, where ${m} \in \mathcal{M}_u$ (i.e., a universal set of messages) is a message, and ${d} \in \{input,output\}$ specifies whether a message is consumed (i.e., it is an input message) or produced (i.e., it is an output message). A message can have a payload, which is a set of values conveyed by the message.
\end{definition}

\begin{table*}[!ht]
 \begin{threeparttable}
\centering
	\caption{Helper functions} 
	\label{tab:helperfunction}
    \begin{tabular}{|p{2.3cm}|p{15.3cm}|}
    \hline
     \small \textbf{Function} &      \small \textbf{Description}  \\ 
    \hline

    \small \codee{inp($c$)}   &\small  returns possible input messages of component ${c}$. E.g., \codee{inp(UC)=\;\{on$,$ \,off\}}.  \\
     \hline
    \small \codee{in\_t($s$)} &  \small  returns incoming transitions to state  $s$. E.g., \codee{in\_t($en_{1}$)=\;\{$t_{12}$\}}. \\
         \hline
    \small \codee{out\_t($s$)}    & \small  returns outgoing transitions from state ${s}$. E.g., \codee{out\_t($en_{1}$)=\;\{$t_{21}$\}}.\\
     \hline
   \small  \codee{handled($s$)}    & \small returns triggers of outgoing transitions  of  ${s}$ and \codee{parents($s$)}. E.g., \codee{handled($s_{11}$)=\;\{on\}}.  \\
    \hline
    \small \codee{root($sm$)}   & \small  returns root of the HSM  ${sm}$.  \\
    \hline
   \small \codee{child($s$)}    & \small  returns states inside  state  ${s}$. E.g., \codee{child(root(CTRSM))=\;\{$s_{11},s_{21},s_{22},s_{23},en_{1},in_{11},c_{11}$\}}. \\
    \hline

   \small \codee{parent($s$)}     & \small  returns the first-level container state of state ${s}$. E.g.,  \codee{parent($s_{21}$)=\;\{$c_{11}$\}}. \\
     \hline
   \small \codee{parents($s$)}     & \small returns all container states of state ${s}$. E.g.,  \codee{parents($s_{21}$)=\;\{$c_{11},$ root(CTRSM)\}}. \\
     \hline
    \codee{deadlock($s$)}   & \small  returns $true$ if state \codee{$s$} and its parents do not handle any message (i.e., \codee{handled(s)=$\emptyset$}). E.g., \codee{deadlock($s_{23}$)= true} and \codee{deadlock($s_{11}$)= false}.   \\
     \hline
    \codee{u\_h($s,h$)} & \small if $parent(s) \centernot= \centernot \emptyset$ and $s \in \mathcal{S}_b$, updates the last visited state of  $parent(s)$ to $s$ (i.e., entry of $parent(s)$ in $h$) and returns the updated $h$. \\
     \hline
    \codee{head($q$)}  & \small  reads, removes, and returns the first element in queue $q$.   \\
     \hline  
    \codee{next\_s($s_c, \mathcal{H}$)}  & \small  (1) returns the last visited state inside state $s_c$ from history $\mathcal{H}$, (2) if (1) is unsuccessful (i.e., the composite state is active for the first time), returns the default state (initial state) inside $s_c$, and (3) if (1) and (2) are unsuccessful, returns $\emptyset$. 
    \\
     \hline
    \codee{next\_t($s,\mu$)} & \small  checks state $s$ and its ancestors in bottom-up order, and returns the first (i.e., most deeply nested) outgoing transition, which can be triggered by message $\mu$. It returns  $\emptyset$ if no transition can be triggered. E.g., \codee{next\_t($s_{21},$ on)= $\emptyset,$} \codee{next\_t($s_{21},$ timeout)= $t_{22}$}.\\
     \hline
    \codee{up\_s($s,t$)}  & \small  returns $s$ and a subset of its parents in bottom-up order from state $s$ to the state that $t$ originated from. E.g., \codee{up\_s($s_{21},t_{13}$)= $\{s_{21}, c_{11}\}$}. \\
    \hline
    \codee{eval($\mathcal{E},g$)}   & \small  evaluates guard $g$ based on the values in map $\mathcal{E}$ and returns the result. \\
    \hline
    \codee{exec($\mathcal{E},a_1 \dots a_n$)}   & \small executes a sequence of actions  $a_1 \dots a_n$ based on the values in map $\mathcal{E}$ and returns the updated $\mathcal{E}$. \\
    \hline
    \end{tabular}
    \begin{tablenotes}
      \small
      \item  	( \small $s$ is a state, $sm$ is an HSM, $t$ is a transition, $q$ is a queue, $\mathcal{E}$ is mapping from variables to their values, $s_c$ is a composite state, $a_1 \dots a_n$ is a sequence of actions. )
\end{tablenotes}

\end{threeparttable}
    \bigskip 
\end{table*}

\begin{definition} \textit{(Component)}  \label{def:component}
 Let us define a component as a tuple $\langle \mathcal{P},\,\mathcal{V},\, \beta\rangle$, where $\mathcal{P} \subseteq \mathcal{P}_u$ (i.e., a universal set of ports) is a set of ports, $\mathcal{V}$ is a set of variables, and $\beta$ refers to the specification of the component's behavior. A port is defined as a pair $({t},{conjugated})$, where ${t}$ denotes the type of the port where the type of a port is an interface, and ${conjugated} \in$ \codee{\{true, false\}} specifies whether or not the port is conjugated. The direction of messages of conjugated ports is reversed.  We will use conjugation to ensure that connected ports are compatible by requiring that (1) they have the same type, and (2) one of them is conjugated while the other one is not. Non conjugated ports are also called base ports.
\end{definition}

 \begin{definition} \textit{(Structure of an RTE system)}  \label{def:system}
 Let us define the structure of an RTE system as a tuple $\langle \mathcal{C},\mathcal{I},{con},in\rangle$, where $\mathcal{C}$ is a set of components,  $\mathcal{I}$ is a set of interfaces,  ${con}$ is a connectivity relationship $\subseteq \mathcal{P}_u \times \mathcal{P}_u$, and $in$ is an acyclic containment relation $\subseteq \mathcal{C} \times \mathcal{C}$. Whenever two ports $p1, p2$ are connected by $con$ (i.e., $(p1,p2) \in con$) then both have the same type (i.e., $p1.t=p2.t$) and exactly one of them must be conjugated. This condition ensures that connected ports are \lq compatible\rq{}. 

Often, MDE tools provide timing services that can be used to define timed behaviors. To support time in our formalization, we assume that an RTE system contains a \codee{timing} interface \codee{\{(startTimer, input)}, \codee{(timeout, output)\}} and a component called \codee{RTS} with a port of type \codee{timing}. Any component using timing services requires a connection with the \codee{RTS} component. 
\end{definition}

Let us exemplify the above definition in the context of the running example (see Fig.~\ref{fig:abmstructure}). The \textit{CTR} is connected to \textit{UC} and \textit{SLD} using two ports \textit{UCPort} (base port) and \textit{SLDPort} (conjugate port), which are typed by interfaces \textit{ControlP} and \textit{StopLightP}, respectively. \textit{ControlP} has two messages (\textit{on()} and \textit{off()}) and \textit{StopLightP} has five messages (\textit{red(), green(), yellow(), on()}, and \textit{off()}).

\begin{definition}\textit{(Action language)} \label{def:actionlang}
Action languages support primitive operations such as accessing/updating variables, arithmetic/logical expressions, control flow constructs, and sending messages. MDE tools provide action languages either by adapting a subset of well-known programming languages or by offering a specific, dedicated action language. E.g., Papyrus-RT uses a subset of C++ as the action language, UML assumes the use of the UML \codee{Alf} action language~\cite{Alf}, and \codee{YAKINDU}~\cite{YAKINDU} provides its own action language. In this work, we assume the existence of an action language with the standard capabilities, but not define a particular syntax for it.
\end{definition}

\begin{definition} \textit{(Hierarchical State Machine (HSM))}  \label{def:usm}
We specify the behavior of a component ${c}$ using a hierarchical state machine (\codee{HSM}) that is defined as a tuple  $\langle \mathcal{S},\mathcal{T}, {in} \rangle$.  $\mathcal{S} = \mathcal{S}_{b}\cup \mathcal{S}_{c} \cup \mathcal{S}_{p}$ is a set of states,  $\mathcal{T}$ is a set of transitions, and ${in} \subseteq \mathcal{S}_{c} \times  (\mathcal{S} \cup \mathcal{T})  $ denotes an acyclic containment relation. 
States can be basic ($\mathcal{S}_{b}$), composite ($\mathcal{S}_{c}$), or pseudo-states ($\mathcal{S}_{p}$). Basic states are
primitive states that the execution stays in until an outgoing transition is triggered. Composite states encapsulate a sub-state machine. Pseudo-states are transient control-flow states. There are six kinds of pseudo-states, called $initial$, $choice$-$point$, $history$, $junction$-$point$, $entry$-$point$, and $exit$-$point$, (i.e.,  $\mathcal{S}_{p} = \mathcal{S}_{in} \cup \mathcal{S}_{ch} \cup \mathcal{S}_{h} \cup \mathcal{S}_{j} \cup \mathcal{S}_{en} \cup \mathcal{S}_{ex}$). Composite and basic states can have entry and exit actions that are coded using an action language. 
\end{definition}

\begin{definition} \textit{(Transition)}
Let $inp(c)$ refer to the messages that can be received by component $c$. 
A transition ${t}$ is a 5-tuple ${(src, guard, trig, act, des)}$, where  ${src, des} \in S$ refer to non-empty source and destination of the transition respectively,  ${guard}$  is a logical expression coded using the action language,  ${trig} \subseteq {inp(c)}$ is a set of messages that trigger the transition, and $act$ is the transition's action coded using the action language. 
\end{definition}

Figure~\ref{fig:CTRHSM} shows an example of \codee{HSM}, and the corresponding graphical notations.

\begin{definition} \textit{(Helper functions)}
Table~\ref{tab:helperfunction}  lists the helper functions (along with samples in the context of the running example, if possible) that will be used in the rest of the paper. Note that we treat the root of an \codee{HSM} as a composite state, which can be accessed using the \codee{root(HSM)} function. 


\end{definition}
\begin{definition} \label{def:weelformed}\textit{(Well-formedness constraints of \codee{HSM}s)}
Following~\cite{Ernesto,RSART, PapyrusRT}, we define the well-formedness constraints of \codee{HSM}s as follows:
\begin{itemize}
    \item Only transitions that start from a choice-point can have a guard, and no transition that starts from a pseudo-state can have a trigger. This constraint is defined to simplify the formalization, and our implementation addresses this case. 
    \item There are no AND-states (orthogonal regions), and UML concepts \textit{fork}, \textit{join}, \textit{shallow history}, and \textit{final states} are also not used.
    \item Transitions cannot cross state boundaries, i.e., $\forall t \in \mathcal{T} :$ \codee{parent(t.src)=parent(t.des)}. Entry-point and exit-point states can be used to create transitions with different parents. 
    \item States do not have idle (\textit{do}) actions.
    \item There is no notation for history. Instead, any transition to a composite state is assumed to end in an implicit history state inside the composite state.
    \item Triggers of transitions starting from the same basic or composite state must be disjoint, i.e., $\forall \, t1,t2 \in \mathcal{T}: t1.src=t2.src \wedge t1.src \centernot \in \mathcal{S}_p \implies t1.trig \cap t2.trig=\emptyset$.
    \item None of the pseudo-states except choice-points can have more than one outgoing transition.
    \item Composite states and the root of the \codee{HSM} cannot have more than one initial state.
\end{itemize} 
\end{definition}

Note that except the first constraint, these constraints are also enforced by existing UML-RT tools and none of them has been defined specifically for this study. Still, our approach can be extended to support AND-states and other concepts not offered in UML-RT.



\begin{definition} \textit{(Configuration)} \label{sec:config}
A configuration $\gamma$ of component $c$ is defined as a tuple $\langle \sigma, {\mathcal{E}}, \mathcal{H} \rangle$ where $\sigma \in \mathcal{S}$ refers to the current state of the configuration, ${\mathcal{E}}$ refers to a mapping from the component variables to values, and $\mathcal{H}$ is a partial mapping from composite states to their last visited sub-states. 
\end{definition}

\begin{definition} \textit{(Execution of  HSMs)} \label{sec:semantic}
We use Labeled Transition Systems (LTS) to define the execution semantics of an \codee{HSM} of a component $c$. An LTS is a tuple $\langle \Gamma, \mathcal{A}, \gamma_{0}, \mathcal{Q}, \rightarrow \rangle$, where $\Gamma$ is a set of configurations, $\mathcal{A} $ is the set of actions (i.e., entry, exit, and transition actions defined in \codee{HSM}), $\mathcal{Q}$ is a first-in, first-out (FIFO) queue that stores received messages, $\rightarrow$ is a transition relation (to avoid confusion with the syntax of \codee{HSM}, we use the term \lq execution step\rq{} instead of \lq transition\rq{} in the rest of the paper), and $\gamma_0 \in \Gamma$ is the initial configuration. 
\end{definition}

\begin{definition} \textit{(Execution Step)} 
An execution step is defined as a tuple $\langle \gamma, a_1 \dots a_n, \gamma^\prime \rangle$ that moves the execution from configuration $\gamma=\langle \sigma, {\mathcal{E}}, \mathcal{H} \rangle$ (source configuration)  to configuration $\gamma^\prime= \langle \sigma^\prime, {\mathcal{E}}^\prime, \mathcal{H}^\prime \rangle$ (target configuration), while executing a possibly empty sequence of actions $a_1 \dots a_n \,$ with $a_i \in \mathcal{A}$ for all 1 $\leq i \leq n$
that may result in updating $\mathcal{H}$ and $\mathcal{E}$, and producing outputs. We use the following notation to show an execution step.
\begin{align*}
\langle \sigma, {\mathcal{E}}, \mathcal{H} \rangle \xrightarrow[]{{\mathcal{E}}^\prime \gets exec({\mathcal{E}},a_1 \cdots a_n)} \langle \sigma^\prime, {\mathcal{E}}^\prime, \mathcal{H}^\prime \rangle
\end{align*}
\end{definition}

\begin{definition} \textit{(Stuck Configuration)}
A stuck configuration is a configuration that no execution step can start from, i.e., the execution cannot progress anymore when it reaches a stuck configuration. We use notation $\gamma_{s}\nrightarrow $ to show that configuration $\gamma_{s}$ is a stuck configuration.
\end{definition}
\begin{figure*}[!t]
\[
  \begin{array}{cccc}
    \dfrac{\sigma \in  \mathcal{S}_{p} \setminus (\mathcal{S}_{h} \cup \mathcal{S}_{ch}), \, t = out\_t(\sigma)}{\langle \sigma, {\mathcal{E}}, \mathcal{H} \rangle \xrightarrow[]{{\mathcal{E}}^\prime \gets exec({\mathcal{E}},act(t),\, entry(t.des))} \langle t.des, {\mathcal{E}}^\prime, u\_h(t.des,\mathcal{H}) \rangle} \; \rlap{\text{(1)}} 
   \quad \quad \quad
    \dfrac{ \sigma \in  \mathcal{S}_{p} \setminus (\mathcal{S}_{h} \cup \mathcal{S}_{ch})  ,  out\_t(\sigma) = \emptyset } 
     {\langle \sigma, {\mathcal{E}}, \mathcal{H} \rangle  \nrightarrow  }\;{(2)}
     
     \\
     \\
    \quad
     \dfrac{\sigma \in  \mathcal{S}_{b}  \, , deadlock(\sigma)} 
    {\langle \sigma, {\mathcal{E}}, \mathcal{H} \rangle  \nrightarrow  }\;{(3)}
     
    \quad
    \dfrac{  \mathcal{Q} \centernot = \emptyset ,  \sigma \in  \mathcal{S}_{b} ,\, \neg deadlock(\sigma) \, ,  t = next\_t(\sigma,head(\mathcal{Q})) } 
    {\langle \sigma, {\mathcal{E}}, \mathcal{H} \rangle \xrightarrow[ ]{{\mathcal{E}}^\prime \gets exec({\mathcal{E}}, exit(t.src),exit(up\_s(\gamma.\sigma, t)), act(t),\, entry(t.des))} \langle t.des, {\mathcal{E}}^\prime, u\_h(t.des,\mathcal{H}) \rangle} \; {(4)}
   \quad
    
    \\
    \\
     
    \dfrac{  \mathcal{Q} \centernot = \emptyset ,  \sigma \in  \mathcal{S}_{b}   \, , \neg deadlock(\sigma), next\_t(\sigma,head(\mathcal{Q})) = \emptyset } 
    {\langle \sigma, {\mathcal{E}}, \mathcal{H} \rangle  \nrightarrow} \; {(5)}

   \\
    \\
    \dfrac
      {\sigma \in \mathcal{S}_{c} , s = next\_s(\sigma, \mathcal{H})  }
      {\langle \sigma, {\mathcal{E}}, \mathcal{H} \rangle  \xrightarrow[]{ {\mathcal{E}}^\prime \gets exec({\mathcal{E}},entry(s))} \langle s, {\mathcal{E}}^\prime, u\_h(s,\mathcal{H}) \rangle }\;{(6)}
    \quad
    \dfrac
      {\sigma \in \mathcal{S}_{c} ,  next\_s(\sigma, \mathcal{H})=\emptyset}
      {\langle \sigma, {\mathcal{E}}, \mathcal{H} \rangle  \nrightarrow}\;{(7)}
    \\
    \\
    \dfrac
    {\sigma \in \mathcal{S}_{ch} ,  t = out\_t(\sigma)  \wedge eval(\mathcal{E},t.guard)}
    {\langle \sigma, {\mathcal{E}}, \mathcal{H} \rangle  \xrightarrow[]{{\mathcal{E}}^\prime \gets exec({\mathcal{E}},act(t),\, entry(t.des))} \langle t.des, {\mathcal{E}}, u\_h(t.des,\mathcal{H}) \rangle }\;{(8)} 
    \quad
   \dfrac
      {\sigma \in \mathcal{S}_{ch} , \centernot \exists t \in out\_t(\sigma) |\, eval(\mathcal{E},t.guard)}
      {\langle \sigma, {\mathcal{E}}, \mathcal{H} \rangle  \nrightarrow}\;{(9)} 
  \end{array}
\]
\caption{Execution rules of an \codee{HSM} adapted from ~\cite{Ernesto,von2006formal}}
\label{fig:semantics}
\end{figure*}
\begin{definition} \textit{(Initial Configuration)} \label{def:initconf}
The initial configuration of an $HSM$  is is defined as $\gamma_0=\langle initial, {\mathcal{E}_0}, \emptyset \rangle$, where $initial$ refers to the initial state inside the root of the \codee{HSM} (i.e., $initial = \mathcal{S}_{in} \cap child(root(HSM))$) and ${\mathcal{E}_0}$ refers to default values of the variables. The execution of the $HSM$ starts from its initial configuration and if the initial state of the $HSM$ is not defined, the execution cannot start (missing initial state).
\end{definition}

\begin{definition} \textit{(Execution Rules of an HSM)} \label{def:rules}
Let us assume that $\gamma= \langle \sigma, {\mathcal{E}}, \mathcal{H} \rangle$ refers to the current configuration. The rules in Figure~\ref{fig:semantics} define the operational semantics\cite{plotkin1981structural} of \codee{HSM}s. The presentation of the rules makes use of definitions from Table~\ref{tab:helperfunction}. The rules are adapted from the execution semantics of UML-RT, presented in~\cite{Ernesto,von2006formal}. 

\noindent \textbf{Rule-1, 2:} These rules are applicable to configurations whose current state is one of the pseudo-states, except for history and choice-point. According to Rule-1, an execution step is taken if there is an outgoing transition from the current state that executes the related actions and moves the execution to a new configuration. Conversely (Rule-2), if there is no outgoing transition, the execution stops there, and the current configuration is considered stuck (issue \rq broken chain\lq{} in 
Sec.~\ref{sec:staticanalysis}). \\
\textbf{Rule-3, 4, 5:} These rules are applicable to configurations whose current state is a basic state. 
If the current state is a deadlock state (Rule-3), the execution stops there, and the current configuration is considered stuck (issue deadlock state). Otherwise, if a message exists in the queue, one of the following rules is applied based on the result of the function $next\_t(\sigma, head(\mathcal{Q}))$ (Ref. Table~\ref{tab:helperfunction}), a transition can be triggered, which results in an execution step that executes the related actions and moves the execution to a new configuration as shown in the bottom of the rule (Rule-4). 
Conversely (Rule-5), if a transition cannot be triggered (i.e., the incoming message is an unexpected message), an execution step cannot be taken. We consider the configuration to be stuck. However, it is also possible to configure the RTS to throw away the unexpected messages. As a result, the execution can recover and continue. We argue that in the domain of RTE systems, in which most of the applications are safety-critical, it is not safe to throw away any message. 

\noindent \textbf{Rule-6, 7:} These rules are applicable to configurations whose current state is a composite state (implicit history state). If function $next\_s(\sigma,\mathcal{H})$ (Ref. Table \ref{tab:helperfunction}) returns a state, then an execution step is taken that applies the entry code of the related composite state and moves the execution to a new configuration, as shown in the bottom of the rule (Rule-6). Conversely, if the selection is unsuccessful, the execution cannot move, and the configuration is a stuck configuration (Rule-7). This can happen due to two reasons: (1) the current state has no child (issue childless composite state), (2) the current state has no initial state (issue missing initial state). 

\noindent\textbf{Rule-8, 9:} These rules are applicable to configurations whose current state is a choice-point. Guards of the outgoing transitions from the current state are evaluated, and the first transition whose guard evaluates to \codee{true} is selected. This results in an execution step that executes the related action code and moves the execution to a new configuration, as shown in the bottom of the rule (Rule-8). Conversely, if none of the outgoing transitions' guards holds (issue non-exhaustive guards), the execution cannot move, and the configuration is a stuck configuration. 

Note that Rules~2, 3, 5, 7, and 9 (all of which are related to stuck configurations) can be merged into one rule. However, we use the different rules for the sake of clarity. 
\end{definition}

\begin{definition} \textit{(Execution of an RTE system)}
The execution of an RTE system can be defined as a collection of its components' \codee{HSM} executions, which interact with each other by passing messages. We do not describe the details of the composition here, and we assume that the RTE system execution is managed by a controller. The controller is responsible for scheduling and message-passing between components, and guarantees that an incoming message will be fully processed before the processing of the next message starts (run-to-completion semantics).

\end{definition}

\newtheorem{prop}{Proposition}
\newcommand{\RNum}[1]{\uppercase\expandafter{\romannumeral #1\relax}}
\newtheorem{lemma}[theorem]{Lemma}

\section{A Conceptual Framework} \label{sec:framework}
\begin{figure*}[t!]
    \centering
    \includegraphics{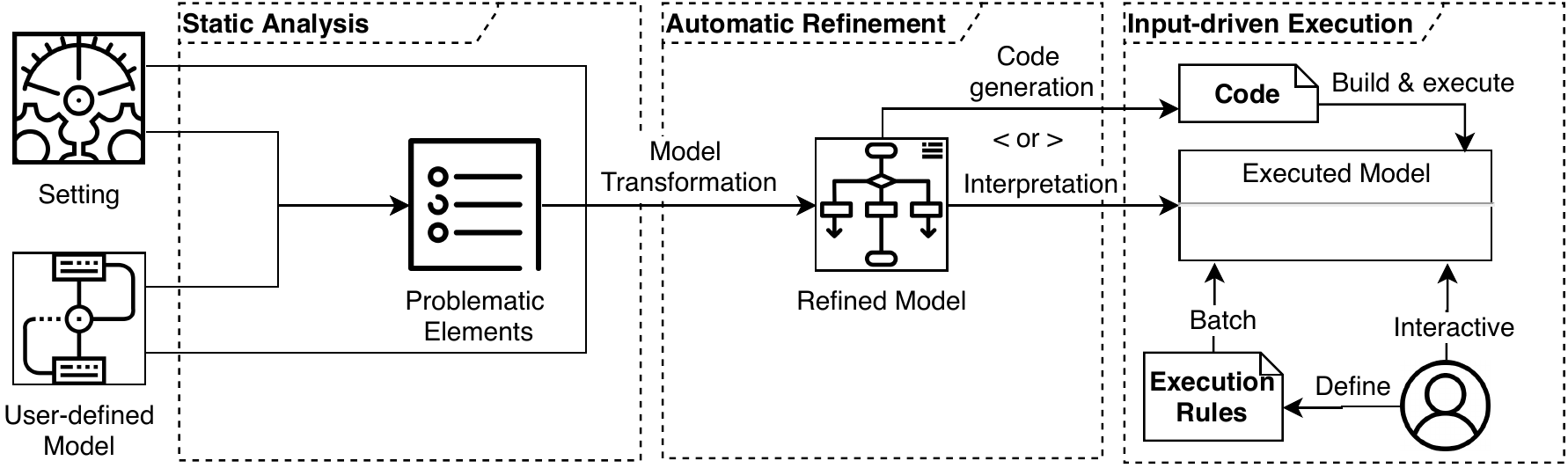}
    \caption{A Conceptual Framework for Executing Partial Models}
    \label{fig:overview}
\end{figure*}

Figure~\ref{fig:overview} shows a conceptual framework for executing partial models, which consists of three parts (\textit{Static Analysis}, \textit{Automatic Refinement} and \textit{Input-driven Execution}). In the following, we discuss the setting and the three parts.
\subsection{Setting} As discussed in Section~\ref{sec:introduction}, a partial model can be executed for different purposes. In all cases, the completeness level of a model is often decided by different stakeholders, based on different constraints and goals. We do not have an automatic check to determine when a model is complete. Instead, we allow users to specify the completeness level of each component based on their need. Supported levels are $clevels=$\textit{\{complete, partial, absent/ignored\}}. \\
\textit{Complete} components are assumed to be complete, and are not required to be analyzed and refined. { By default, each component is assumed to be complete, unless its completeness level is explicitly set to something else. }\\
\textit{Partial} components are assumed to be incomplete. Thus, their current specification (structure and behavior) is analyzed and refined.\\ \textit{Absent/ignored} components are assumed to have no behavior specification. However, their existence may be necessary for the execution of other (\textit{partial} and \textit{complete}) components, due to the dependency between them. Thus, the \textit{absent/ignored} components are analyzed based on their structure (inputs and outputs) and possible dependencies of other components on them. Then they are given behaviour sufficient for simulation. 

The setting allows the execution of the models for different purposes. E.g., in the context of the running example which is discussed in Subsection~\ref{sec:runningexample}, we can set the completeness level of \textit{CTR} to \textit{partial}, that of \textit{SLD} to \textit{complete}, and that of \textit{UC} to \textit{absent/ignored} to execute the system for Scenario-1 (evaluation of design decision). For scenario-2 (unit testing), the level of \textit{CTR} should be set to \textit{partial} and the levels of the other components to \textit{absent/ignored}. 


\subsection{Static Analysis} Assuming that the execution semantics of the language is defined, we perform static analysis with respect to the execution semantics to detect the problematic elements that can cause a problem for the execution. Depending on the semantics of the language different types of problems may be detected. In the context of state machines, the problems associated with executing partial models fall into two groups: \textit{lack of progress} and \textit{lack of reachability}. The former is related to situations in which the execution cannot progress anymore from a certain point. The latter concerns the execution being unable to reach certain, specific states. The static analysis is performed on the user-defined model and identifies problematic elements including: (1) missing elements, (2) existing elements with problematic specifications, and (3) missing and unhandled inputs. 

\subsection{Automatic Refinement} During the refinement phase, depending on the results of the static analysis and the setting, the user-defined model is refined automatically using model-to-model transformation techniques. The goal of the refinement is to fix the problematic elements or modify them in such a way that users can provide more information about them during execution. Depending on the modelling language, certain language constructs can be used to enable models to interact with users during the execution. E.g., for \textit{HSM}s, we use choice-points with certain actions and guards. 

The refinement should meet the following constraints: (1) Refined models must preserve the original behavior of the user-defined models. (2) The execution of the refined model must not get stuck, assuming proper inputs are provided. (3) The execution of the refined model must be able to reach all defined states in a finite number of execution steps, assuming proper inputs are provided. (4) The execution of the refined model must allow users to select one of the possible options to fix the problematic element. The options must be exhaustive and include all possible situations which can be applied at design time without limiting the users to only a subset of them. 

\subsection{Input-driven Execution}
The refined model can be executed via interpretation or code generation. During execution, the executed model provides an interface for reading user input either in interactive or batch mode. It is also crucial to provide debugging facilities. Thus, users can investigate the execution before providing inputs.

\section{Application of the Framework to Partial UML-RT Models} \label{sec:application}
Generally, the discussed framework to allow execution of partial models can be applied in the context of different modeling languages. However, since the execution of models is a language-dependent concept, there is no way to provide a generic implementation of the framework using existing techniques and tools. In the rest of this section, we discuss the application of the framework for executing partial UML-RT models and demonstrate a tool that embodies the framework. 

\subsection{Static Analysis} \label{sec:staticanalysis}
In the following,  we discuss the details of the static analysis of UML-RT models, with respect to the execution semantics of UML-RT as discussed in Def.~\ref{sec:semantic}. We categorize the problems based their effect on lack of progress or reachability. Note that  a lack of progress entails a lack of reachability. \\ 

\subsubsection{{Lack of progress}} Based on the execution rules of \textit{HSM}s (see Def.~\ref{def:rules}), the execution of an \textit{HSM} can be stopped due to several issues, which can be divided into two groups, as follows. 

      \textbf{Missing/problematic elements:}  The execution of an \textit{HSM} cannot start, or moves to a stuck configuration, due to the following issues. 
      \begin{itemize} 
      \item \textit{P1}: missing initial state (see Def.~\ref{def:initconf}), 
      \item \textit{P2}: childless composite states (see Rule-$7$ of Fig.~\ref{fig:semantics}), \item \textit{P3}: broken chain  (see Rule-$2$ of Fig.~\ref{fig:semantics}), 
      \item \textit{P4}: deadlock state  (see Rule-$3$ of Fig.~\ref{fig:semantics}) , 
      \item \textit{P5}: unexpected messages  (see Rule-$5$ of Fig.~\ref{fig:semantics}),
      \item \textit{P6} : non-exhaustive guards of choice-points (seeRule-$9$ of Fig.~\ref{fig:semantics}). 
      \end{itemize}
      Except for P6, the elements with these issues can be described by queries on the structure of the \textit{HSM}, as follows.  
\begin{align*}
\begin{tabular}{ll}
    \textit{P1} $\leftarrow$ & $\{s_c \in (root(HSM) \cup  \mathcal{S}_c):\; \mathcal{S}_{in} \cap child(s_c) =\emptyset \}$ \\ 
    \textit{P2} $\leftarrow$ & $\{s_c \in (root(HSM) \cup  \mathcal{S}_c):\; child(s_c) =\emptyset \}$ \\ 
    \textit{P3} $\leftarrow$ & $\{ s \in \mathcal{S}_{p} \setminus  (\mathcal{S}_{h} \cup \mathcal{S}_{ch}): \;  out\_t(s)=\emptyset\} $ \\
    \textit{P4} $\leftarrow$ & $\{ s \in \mathcal{S}_b: \;   handled(s)=\emptyset \} $ \\ 
    \textit{P5} $\leftarrow$  & $\{ s \in \mathcal{S}_{b} : \; inp(c) \, \setminus  handled(s) \centernot = \emptyset \} $
\end{tabular}
\end{align*}

As for \textit{P6}, we assume that all choice-points have this problem, (i.e., \textit{P6} $\leftarrow$  $\{ s \in \mathcal{S}_{ch}$\}). This is an overestimation, and covers all possible situations of \textit{P6}. Checking the exhaustiveness of guards of a choice-point during design time is a difficult problem, and requires expensive computation. Thus, fixing this problem at design-time can increase the refinement time significantly, and even make it unsolvable. Also, the applicability of the existing techniques on partial models is not supported by default, and requires extra work and research. 
\newcommand{\hrulealg}[0]{\vspace{1mm} \hrule \vspace{1mm}}
\begin{algorithm}[t!]
\SetKwInOut{Input}{Input}
\SetKwInOut{Output}{Output}
\Input{A UML-RT model \textit{sys} and a setting \textit{conf} }
\Output{A refined model}
\SetAlFnt{\scriptsize\sf}
\SetKwProg{Fn}{Function}{}{}
\DontPrintSemicolon
{
\hrulealg
Add a debugging interface \textit{dbg\_int} and a debugging agent \textit{dbg\_agent} into \textit{sys} \\
Add ports types with \textit{timing} and \textit{dbg\_int} into \textit{dbg\_agent}
\ForAll(\tcp*[f]{sys.$\mathcal{C}$ (components of sys)}){$c \in$ \textit{sys}.C}{ 
    \Switch(\tcp*[f]{{c.conf} (setting of c)}){\textit{c.conf}}{ 
        \Case{\textit{partial}}{
            Add port $p$ of type  \textit{dbg\_int} into component $c$ \\
            Add a connection using port $p$ with \textit{dbg\_agent} \\
            $c.\beta \gets$ refineHSM$(c.\beta,\, c)$
        }
        \Case{\textit{absent/ignored}}{
            Add port $p$ of type  \textit{dbg\_int} into component $c$ \\
            Add a connection using port $p$ with \textit{dbg\_agent} \\
            Delete all elements from HSM of $c$ ($c.\beta$) \\
            $c.\beta \gets$ refineHSM$(c.\beta,\, c)$
        }
        \Other{
            \tcp{No refinement}
        } 
    }
    } 
    \textit{dbg\_agent}$.\beta \gets$ refineHSM$( \textit{dbg\_agent}.\beta,\, c)$
}

\caption{Refinement of a Partial UML-RT Model}\label{refinemodel}
\end{algorithm}

 \textbf{Missing inputs:} A prerequisite for taking an execution step from basic states is the reception of a new message (see Rule-4 of Fig.~\ref{fig:semantics}) that can enable an outgoing transition from the current state. The execution can be stuck, if the required messages for triggering possible transitions are not produced by the connected components. This can happen for two reasons: (1) the connected component lacks a behavior specification, i.e., components are set as \textit{absent/ignored}  (\textit{P7}), (2) the behavior of a connected component is partial (\textit{P8}). Detecting \textit{P7} is trivial and can be determined from the interface specification of components. Let ${I}$ be a set of possible input messages of all \textit{partial} and \textit{complete} components, and let ${O}$ be a set of possible output messages of \textit{absent/ignored} components, then \textit{P7} $\leftarrow$  ${I}$ $\cap$ ${O}$. \\ 
Detecting \textit{P8} at design time suffers from a similar problem as \textit{P6}. Thus, we overestimate again and assume that all \textit{partial} components have this problem, (i.e., \textit{P8} $\leftarrow$ all \textit{partial} components).  
As we will discuss later, we provide debugging commands for sending messages to the other components from \textit{partial} components, during the execution. That way, users can fix this problem by manually injecting the related messages during the execution.

\subsubsection{Lack of Reachability}  Anything causing the lack of progress problem also causes a lack of reachability. There is no way for the execution to reach any state after being stopped. In addition, the following missing or problematic elements can cause a lack of reachability.  
\begin{itemize}
{
\item \textit{P9}: isolated states are states that do not have incoming transitions (with the exception of initial and history states). There is no way for the execution to reach these states.
}
\item  \textit{P10}: not-takeable transitions originate from a basic or composite state and have no trigger. 
\item 
{
\textit{P11}: as discussed, one of the main benefits of the execution of partial models is enabling early evaluation of different design decisions, the support of which requires all states to be reachable during the execution in a finite number of steps. Otherwise, some design decisions cannot be evaluated due to the lack of reachability issue.  For example, in the context of the running example (see Fig.~\ref{fig:CTRHSM}), assume that transition \textit{t22} is missing, and a user needs to evaluate the effect of action of \textit{t23}. The evaluation is not possible without steering the execution to state \textit{$s22$}. Thus, we define another condition that concerns steering the execution from configurations whose current state is a basic state to any configuration whose current state is any basic state. }
\end{itemize}
Except for \textit{P11}, the elements with these issues can be queried from the structure of a component $c$ with an \textit{HSM} as follows.

\begin{align*}
\begin{tabular}{ll}
    \textit{P9} $\leftarrow$ & $\{  s \in \mathcal{S} \setminus (\mathcal{S}_{in} \cup  \mathcal{S}_{h})  : \;  in\_trans(s)=\emptyset \}$ \\ 
    \textit{P10} $\leftarrow$ & $\{  t \in \mathcal{T}: \; src(t) \in \mathcal{S}_{b} \cup \mathcal{S}_{c} \wedge trig(t)= \emptyset \}$ \\ 
\end{tabular}
\end{align*}

As for \textit{P11}, this capability needs to be addressed in all basic stats, i.e., \textit{P11} $\leftarrow \mathcal{S}_b$.

Note that sets $P1-P11$ are not disjoint and an element may have several issues. For instance, in the context of the running example (see Fig.~\ref{fig:CTRHSM}), $s_{23}$ is a deadlock state and has a reachability problem, i.e., $s_{23} \in P4 \cap P11$. 

\subsection{Refinement of Partial UML-RT Models}
\label{sec:refinment}

\begin{algorithm*}
\SetKwInOut{Input}{Input}
\SetKwInOut{Output}{Output}
\Input{An HSM $sm$ and a component $c$ }
\Output{A refined HSM}
\SetAlFnt{\scriptsize\sf}
\SetKwProg{Fn}{Function}{}{}
\DontPrintSemicolon
{
\hrulealg
\tcp{The following loop refines states in order of their nesting level, with the least deeply nested state (root(HSM)) refined first.}
\ForAll{$s_{c} \in root(sm) \cup ( sm.\mathcal{S} \in  \mathcal{S}_{c})$}{ 
    $dec\_p \gets$ add\_state$(s_{c},\mathcal{S}_{ch})$   \tcp{Add decision point} 
    
    \uIf(\tcp*[h]{Fix childless composite state (P2)}){$s_{c} \in P2$}{
         $state\_p\_h \gets$ add\_state$(s_{c},\mathcal{S}_{b})$ 
    }
    \uIf(\tcp*[h]{Fix missing initial state (P2)}){$s_{c} \in P1 $}{ 
        add\_state$(s_{c},\mathcal{S}_{in})$ 
    }
    \ForAll{$s_{p} \in (child(s_{c}) \cap P3 \setminus dec\_p$)}{
        add\_trans$(s_p,dec\_p)$ \tcp*[l]{Fix broken chain (P3)}
    }
    \ForAll(\tcp*[h]{Fix non-exhaustive guards for choice-points (P6)}){$s_{ch} \in child(s_{c}) \setminus  dec\_p $}{
        $t_1 \gets$ add\_trans$(s_{ch},dec\_p)$ \\
        $t_1.guard \gets \neg \bigvee \big(guard(out\_trans(s_{ch})\big)$ 
    }
    
    \ForAll(\tcp*[h]{Fix unexpected messages (P5), deadlock states (P4), and step 1 of fix for P11}){$s_{b} \in child(s_{c}) \cap P11$}{
        \uIf{$s_{b} \in (P4 \cup P5$)}{
        $t_2 \gets$ add\_trans$(s_{b},dec\_p)$ \\
        $t_2.trig \gets inp(c) \setminus handled(s_{b})$ 
    }
    }
    \ForAll( \tcp*[h]{Fix isolated states (P9) and step 2 of fix for P11}){$s \in (child(s_{c}) \setminus (\mathcal{S}_{in} \cup dec_p) $}{
       $t_3 \gets $add\_trans$(dec\_P,s)$
    }
    
    \ForAll(\tcp*[h]{Fix not-takeable transition (P10)}){$t \in P10$}{
        $t.src=dec\_p$ 
    }
    
    \ForAll (\tcp*[h]{Step 3 of fix for P11}){$s_{cc} \in child(s_{c}) \cap \mathcal{S}_{c}$ }{  
        $ex\_p \gets ex\_p \cup$ add\_state$(s_{cc},\mathcal{S}_{ex})$  \\
        $en\_p \gets en\_p \cup$ add\_state$(s_{cc},\mathcal{S}_{en})$ \\
        $to\_child \gets$ add\_trans$(dec\_p,en\_p \cap child(s_{cc}))$ \\
        $from\_child \gets$ add\_trans$(ex\_p \cap child(s_{cc}) ,dec\_p)$ 
        
    }
    \uIf(\tcp*[h]{Last step of fix for P11} ){$parent(s_{c}) \centernot = \emptyset$}{ 
        $to\_parent \gets $add\_trans$(dec\_p,ex\_p \cap child(s_{cc}) )$ \\
        $from\_parent \gets$ add\_trans$(en\_p \cap child(s_{cc}),dec\_p)$
    }
}

}
\caption{Refinement of \textit{HSM} (refineHSM)}\label{superHSM}
\end{algorithm*}
In the following, we discuss the details of the refinement, applied to fix the problematic elements (\textit{P1-P11}) extracted during the analysis phases. 

\noindent \textbf{Main Loop of the Refinement.}
Algorithm~\ref{refinemodel} shows the main loop of the refinement of a UML-RT model. It takes a UML-RT model and a setting as inputs. A setting is a set of tuples $\langle  c \in \mathcal{C}, clevel \in clevels \rangle$, where $c$ is a component and $clevel$ specifies the level of completeness of the component. The algorithm first adds a new component to the model, called \textit{dbg\_agent} with an empty \textit{HSM}, creates a debugging interface, and adds debugging and timing ports into \textit{dbg\_agent}. \textit{dbg\_agent} is responsible for receiving from external applications and transferring \textit{dbg} messages to \textit{partial} and \textit{absent/ignored} components. After setting up \textit{dbg\_agent}, the algorithm tries to apply certain refinements based on the setting of the components, as follows. 
\begin{enumerate}[leftmargin=*]
    \item For partial components, it adds a debugging port into the components and creates a connection between them and \textit{dbg\_agent}. This allows them to receive the \textit{dbg} message during the execution, which is essential for fixing elements in \textit{P7-P8}. Then it calls the \textit{refineHSM} function, which applies the behavioral refinement to fix the issues (line\# 4-7).
    \item The behavior of absent/ignored components are removed, which results in an empty \textit{HSM}. Then, their empty \textit{HSM} is refined, which results in an \textit{HSM} that can receive and send all possible input and output messages of the component (line\# 8-12).
    \item Finally, the \textit{HSM} of \textit{dbg\_agent} is also refined as a \textit{absent/ignored} component that results in an \textit{HSM} capable of processing debugging commands (line\# 14).

\end{enumerate}


\begin{figure*}[t!]
\centering
\includegraphics[width=16cm]{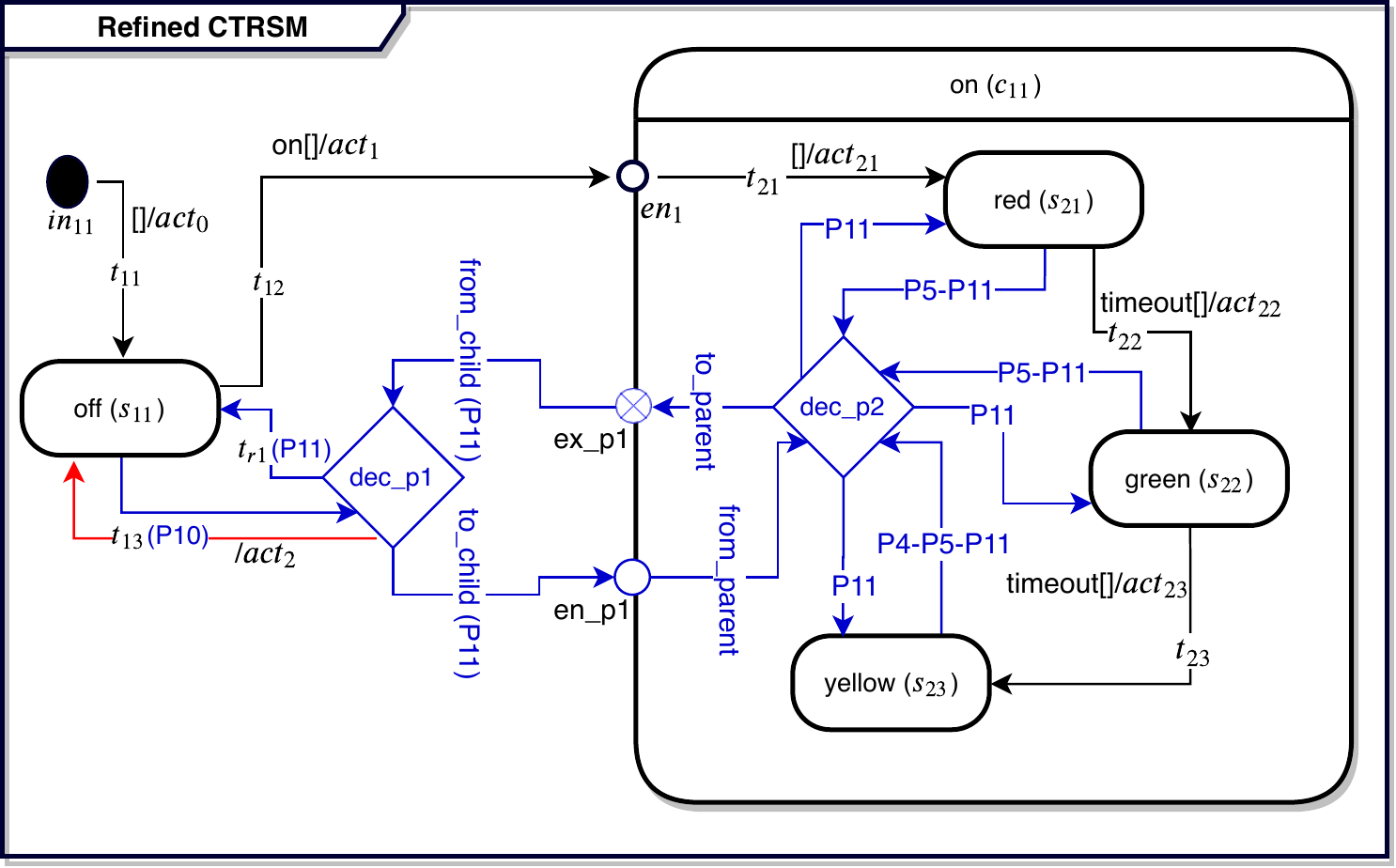}
\caption{Refined version of \textit{CTRSM} in Fig.~\ref{fig:CTRHSM} (added elements are coloured blue, and modified elements are coloured red)}
\label{fig:refinedctr}
\end{figure*}

\noindent \textbf{Behavioral Refinement.} Algorithm~\ref{superHSM} presents function \textit{refineHSM}, which refines the \textit{HSM} of a component with respect to elements in \textit{P1-P11} except for elements in \textit{P7-P8}. Before discussing the details, let us define $add\_state\,(s_c \in \mathcal{S}_{c}, ty \in \mathcal{S}) \rightarrow \mathcal{S}$ as a function that adds a state of type $ty$ inside the $s_c$ and $add\_trans\,(src, trg \in \mathcal{S}) \rightarrow \mathcal{T}$ as a function that adds a transition from state $src$ to state $trg$. The algorithm iterates over all composite states, and the root of the \textit{HSM}, and refines them in $9$ steps, as follows.
\begin{enumerate}[leftmargin=*]
    \item It creates a choice-point state called \textit{dec\_p}. \textit{dec\_p} is used as a decision point during the execution (line\# 2). When a specification is missing, we refine the \textit{HSM} so that the execution is directed to \textit{dec\_p}. 
    \item Fix elements in \textit{P2} which have no child, by adding a new basic state inside the related state (line\# 3-4). Note that the added basic state has issues \textit{P4, P5,} and \textit{P9}, and requires the corresponding fixes.
    \item Fix elements in \textit{P1} which miss initial states, by adding a new initial state inside the related state (line\# 5-6). Note that the added initial state has issue \textit{P3}, and requires the corresponding fixes.
    \item  The elements in \textit{P3} (Broken chain) are fixed by adding a transition from the problematic states to \textit{dec\_p} (line\# 7-8). This ensures that the execution will move to \textit{dec\_p} instead of stopping at the problematic states, and thus users can steer the execution to other states from \textit{dec\_p}. 
    \item Fix elements in \textit{P6} (Non-exhaustive guards) by adding a transition from each choice-point to \textit{dec\_p} so that its guard is set to the negation of the disjunction of the outgoing transitions' guards (line\# 9-11). This ensures that the execution moves to the  \textit{dec\_p} if none of the guards of the outgoing transition holds, instead of stopping there. Arguably, this solution is much cheaper than the design time analysis to detect and fix this issue.
    
    \item During this step, a new transition is added from each basic state to \textit{dec\_p} and its trigger is set to all un-handled messages in the state (line\# 13-15). This not only fixes the elements in \textit{P4} (deadlock state) by adding a transition from them to \textit{dec\_p}, but also (1) allows all un-handled messages to be handled as the new transition's trigger (\textit{P5}), and (2) allows the steering of the execution from any basic state to \textit{dec\_p} which is the first step of the fix for elements in \textit{P11}.
    \item A transition is added from \textit{dec\_p} to all basic states and isolated states (line\#16-17). This fixes issue \textit{P9} (isolated states), and also allows the steering of the execution to any basic state from \textit{dec\_p} which is the second part of the fix for \textit{P11}.
    \item To fix not-takeable transitions \textit{(P9)}, their source is changed to \textit{dec\_p} (line\# 18-19). This allows them to be taken whenever the execution reaches \textit{dec\_p}. Since each state has a transition to \textit{dec\_p} which is added in step 6 with a trigger set to all un-handled messages, the not-takable transitions can be activated by any of the un-handled messages. Note that the \textit{dbg} message, which is added using Algorithm~\ref{refinemodel} is assumed to be an un-handled message. 
    \item At the end, an \textit{exit-point} (line\# 21) and an \textit{entry-point} (line\# 22) states are added to each composite state to allow the execution to be steered from their sub-states to states in their parent state and vice-versa, which is the last part of the fix for elements in \textit{P10}. Two transitions \textit{to\_parent} and \textit{from\_parent} allow the execution to be steered from their sub-states to states in their parent state, and transitions \textit{to\_substates} and \textit{from\_upper} allow the execution to be steered from states in their parent state to their sub-states (line\# 20-27).

\end{enumerate}

Note that the refinement algorithms do not contain details for the actions which are added to \textit{HSM}s. E.g., (1) added actions to the \textit{HSM} of \textit{dbg\_agent} for processing and injecting the message \textit{dbg}, (2) guards of outgoing transitions from decision points which are set in a way that allows users to select one of them, (3) actions of incoming transitions to decision points that call a function to read user input. Interested readers can refer to the source code of the refinement~\cite{pmdebuggersource}.

\subsubsection{Refinement Result on the Running Example}

Figure~\ref{fig:refinedctr} shows the result of running Algorithm~\ref{superHSM} on the partial \textit{CTRSM} with \textit{partial} completeness level in which transitions and states are annotated with corresponding issues \textit{P1-P11}. Let us review some examples of how the execution can be performed despite missing specifications: (1) No transition from \textit{yellow} to \textit{red} was possible in the original model. This is fixed by adding transitions from yellow and red to $dec\_p$ and vice-versa. Thus, any of the input messages or \textit{dbg} messages can move the execution to state \textit{dec\_p} from state \textit{yellow} where users can select one of the outgoing transitions (e.g., the transition from \textit{dec\_p} to state \textit{red}). (2) The transition $t_{13}$  is not-takeable in the original model and its action cannot be executed. In the refined model an \textit{off} or \textit{dbg} message can move the execution to \textit{dec\_p} in which the transition $t_{13}$ is one of the possible transitions and can be selected and its action be executed.


\subsection{Execution of Refined Partial UML-RT Models}
\label{sec:inputdriven}
In this section, we discuss our method for the execution of 
incomplete UML-RT models. The discussion will emphasize key 
concepts over low-level implementation detail. The definitions below identify two such concepts. 
\begin{definition} \textit{(Execution Context)}  \label{def:execcontext}
{Intuitively, an execution context captures the most relevant runtime information of an execution stopped at some decision point. Formally, an execution context is a tuple $\langle \gamma,dec\_p ,m,\mathcal{O} \rangle$, where $dec\_p$ refers to the decision point at which the execution is stopped, $\gamma$ denotes the configuration (see Def.~\ref{sec:config}) right before the execution reached $dec\_p$, $m$ denotes the last processed message by the \textit{HSM} (the trigger of the most recently taken transition starting from $\gamma.\sigma$), and $\mathcal{O}$ is a list of possible options available to continue the execution (i.e., the transitions originating from $dec\_p$ ).}
\end{definition}

\begin{definition} \textit{(Execution Rule)} \label{def:execrule}
{We define an execution rule as a tuple $\langle h, b \rangle$, where $h$ refers to the header and $b$ refers to the body of the rule. A header is a tuple  $\langle name, where, when \rangle$, where \textit{name} refers to the name of the execution rule, \textit{where} refers either to the qualified name of a state (\textit{component.state}), name of a component, *, or \textit{*.state} as shown in Listing~\ref{dbgservices} (Line \#7), and \textit{when} refers to a boolean  condition. A body is a sequence of statements as defined in Listing~\ref{dbgservices}. The semantics and use of execution rules is discussed in Section~\ref{sec:batchexec}. 
}
\end{definition}

\subsubsection{Execution Flow of a Refined Partial Models}
As discussed, the partial models are refined by adding decision points where elements are missing or are partial that allow users to execute the partial models and provide information about the missing or partial element during the execution. This requires a mechanism that (1) enables executed models to obtain user input either in interactive or batch mode, (2) provides debugging features to investigate and modify the execution of the model.

\begin{figure}[t!]
    \centering
    \includegraphics{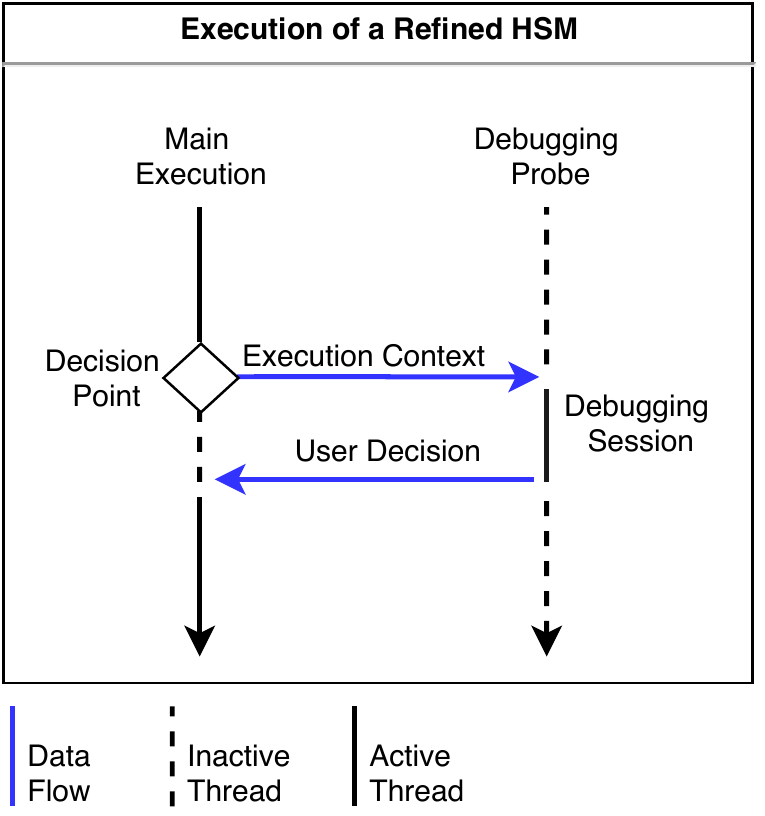}
    \caption{Execution flow of a refined Partial \textit{HSM}}
    \label{fig:execoverview}
\end{figure}

Figure~\ref{fig:execoverview} shows the execution flow of a refined \textit{HSM} in which a debugging probe is hooked into the execution of an \textit{HSM} by adding relevant actions in the initial transition of the \textit{HSM}. When an \textit{HSM} starts executing, two threads \textit{main} and \textit{probe} are started but only one of them is active at each time of the execution. Thread $main$ executes the models as specified until it reaches a decision point where it sends the relevant execution context to the \textit{probe} and waits for the user input. Thread \textit{probe} starts a debugging session (batch or interactive) that allows users to investigate the execution and provide input. At the end of the session, the user input is returned to thread \textit{main} to continue the execution. The interaction between the threads is simply a function call from thread \textit{main} to the \textit{probe} that is implemented using an action in the transitions ending at the decision points. {In the following subsections, we review two execution modes of partial models.}

\label{sec:debuggingprobe}
\lstset{
   language=antlr,
   extendedchars=true,
   basicstyle=\footnotesize\ttfamily,
   showstringspaces=false,
   showspaces=false,
   tabsize=2,
   breaklines=true,
   showtabs=false,
   frame=bt
}
\begin{figure}[t!]

\begin{lstlisting}[caption={High-level Grammar of the Interactive/Batch Execution Commands},captionpos=b,label=dbgservices]
grammar ExecRules;
script: execRule+;
execRule: 'rule' rID=ID 'where' '(' where ')' 
when?
    '{' body=scriptStatement*  '}';
when: 'when' '('guard')';
where: component.state | component| * | *.state;
statement: scriptStatement | interactiveStatement;
scriptStatement: simpleStatement | complexStatement;
interactiveStatement: dbgCmd | umlrtCmd ;
umlrtCmd: sendMsgCmd | replyMsgCmd | receipt;
dbgCmd: viewCmd  | selectCmd | simpleStatement | visited | controlCmd | saveCmd;

\end{lstlisting}
\end{figure}

\subsubsection{Interactive execution}
With the interactive execution, users are allowed to issue debugging commands listed in \textit{interactiveStatement} of Listing~\ref{dbgservices}, e.g., view and modify variables. Most of debugging services are ported from MDebugger~\cite{MDebuggerDemo}. The new debugging commands to facilitate the execution of incomplete models are as follows: 
\begin{enumerate}
\item\textit{viewCmd} lists the possible options to continue the execution. 
\item\textit{selectCmd} allows users to select one of the possible options. {Note that \textit{selectCmd} is the last statement that is applied when the execution has been stopped, and any command after that in the body of the rule is ignored (similar to the \textit{return} statement in many programming languages). Also, \textit{selectCmd} can accept more than one option during the batch execution, and in that case the execution switches to interactive mode to capture the user input.} 
\item\textit{simpleStatement} allows users to define new variables which can be accessed during the debugging session, and to access all attributes (i.e., \textit{HSM}'s variables and newly defined variables during the debugging session) and modify them using arithmetic expressions.
\item{\textit{saveCmd} allows users to save their decision during the interactive session. We discuss this command in detail in Section~\ref{sec:batchexec}}.
\item\textit{umlrtCmd} consists of \textit{send}, \textit{reply}, and \textit{receipt} commands. Commands \textit{send} and \textit{reply} allow users to \textit{send/reply} (inject) messages to other components. Command \textit{receipt} accepts a message as an input and returns \textit{true} if the message is the most recently received message by the component, and \textit{false} otherwise. 
\end{enumerate}

\subsubsection{Batch execution}  \label{sec:batchexec}
Interactive execution stops and delays the execution which is not suitable in some situations, especially for the debugging of time-sensitive systems, to repeat a debugging scenario, or to test and explore a design decision. For these situations, a batch execution mode is supported that allows users to provide inputs using a script that consists of execution rules (see Def.~\ref{def:execrule}). 
An execution rule prescribes how the execution of a refined partial model is to be continued when the current execution context matches the \textit{where} of the rule and the \textit{when} of the rule evaluates to true.  


Depending on the current execution context and defined rules, multiple rules may be applicable at each time, but only one rule can be applied at each time. The rule selection for a decision point of an execution context with component $c$ and the current execution state $s$ is performed by following the steps below:
\begin{enumerate}
    \item Guards of rules whose \textit{where}  (i.e., component and state name) is equal to $c$ and $s$ are evaluated based on the order of their appearance in the script file of the rule. The first rule whose guard holds is selected and applied.
    \item If (1) is unsuccessful, guards of rules whose state name is exactly equal to $s$ and component name is equal to $*$ or empty are evaluated based on the order of their appearance in the rules' script file. The first rule whose guard holds is selected and applied.
    \item If (2) is unsuccessful, guards of rules whose component name is exactly equal to $c$ and state name is equal to $*$ or empty are evaluated based on the order of their appearance in the rules' script file. The first rule whose guard holds is selected and applied.
    \item If (3) is unsuccessful, guards of rules whose state and component name is $*$ are evaluated based on the order of their appearance in the script file of the rule. The first rule whose guard holds is selected and applied.
    \item If none of the above holds, the execution stops and the user is asked to provide input to continue the execution. 
\end{enumerate}

\subsection{Automation} \label{sec:Automation}

In general, the partial models that we consider exhibit one of two kinds of partialness: un-intentional and intentional. 
The former is related to situations (e.g., early debugging) in which the model is still under development, and not yet complete, due to the iterative and incremental nature of the software development process. The latter relates to situations (e.g., unit testing and partial analysis) in which users intentionally execute a complete model as a partial model often by writing scripts for execution rules to mock the ignored or unavailable part of the model. This can, e.g., increase the efficiency of testing or help deal with the unavailability of external components that the model relies on. 

With un-intentional partialness, the user has to provide input (in the form of interactive decisions at runtime or execution rules) to steer the execution of the refined partial model. It is possible that this input resolves the partialness in a generally satisfactory way and that the user then also wants to use it for the next incremental development step and apply it to the design model. However, as we discuss later, our approach is careful to avoid duplicate effort from the user by allowing the input to be used directly to update the design model, instead of requiring the user to provide it again when completing the design model. While this \textit{double effort} may incur negligible costs for the debugging of only a small part of a partial model, its cost can be significant for the debugging of a large part of a partial model, due to the large number of inputs that may need to be provided. 
Note that in the case of intentional partialness, the model is already complete, and writing the scripts for execution rules is only for simulation and mocking of the existing components. 
\begin{figure*}
    \centering
    \includegraphics[width=16cm]{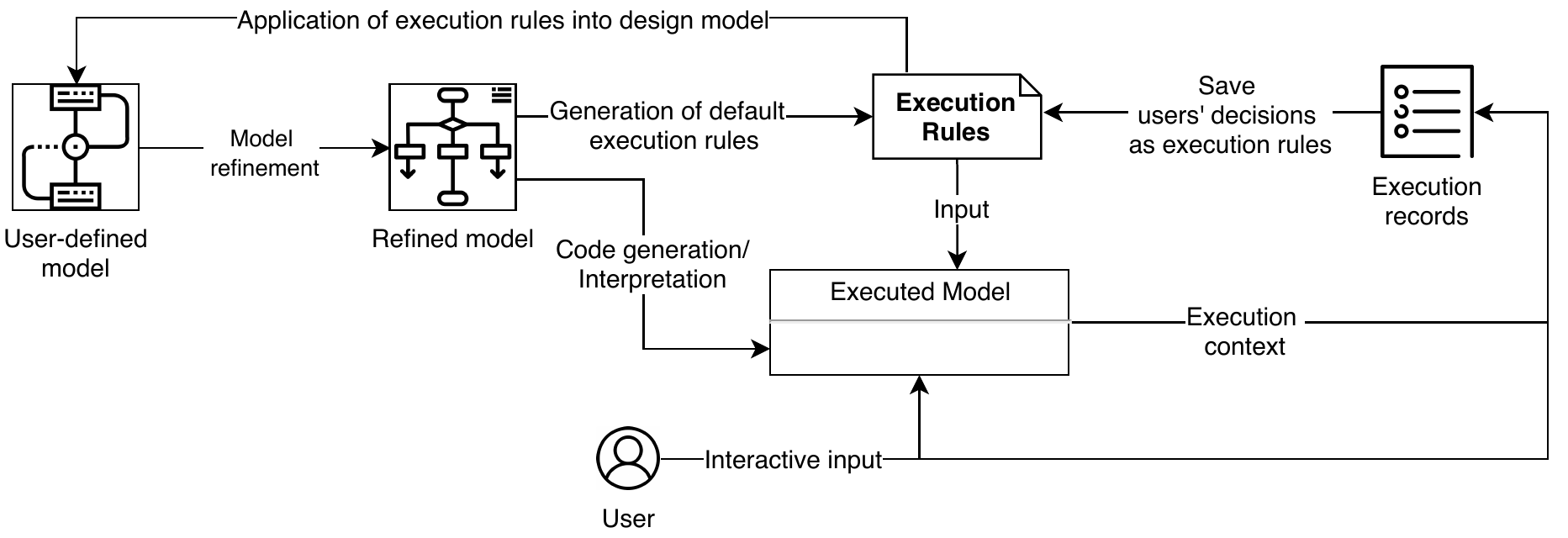}
    \caption{Automated generation of executions rules and completion of the design (user-defined) model}
    \label{fig:automations}
\end{figure*}

To minimize the overhead of writing execution rules, and to reduce the need for double efforts, we provide three automation features, including (1)  generation of default execution rules,  (2) saving of users' interactive decisions as execution rules, and (3) application of execution rules into the design model. As shown in Figure~\ref{fig:automations}, the features are complementary and allow users to automatically create execution scripts and  update the design model by application of the execution rules.
In the following, we discuss the details of each feature.

\label{sec:refinment}

\begin{algorithm}[t]
\SetKwInOut{Input}{Input}
\SetKwInOut{Output}{Output}
\Input{Refined HSM $sm^\prime$, problematic elements (P1-P11)}
\Output{Set of execution rules} 
\SetAlFnt{\scriptsize\sf}
\SetKwProg{Fn}{Function}{}{}
\DontPrintSemicolon
{
\hrulealg
Let $R$ be an empty set \\
Let $D$ be a set that contains all of $dec\_p$ of $sm^\prime$ which have been  added during refinement\\
Let $P\_e$ be exit-point/entry-point states that are added during the refinement \\
\ForAll{$dec\_p \in D $}{ 
   \ForAll{$t \in \{t: t.des=dec\_p \}$}{ 
        \uIf(){$t.src \centernot \in P\_e$}{
        $M \gets  t.trig \setminus dbg$ \\
        \uIf(){$M = \emptyset$} {
            Add message $dummy$  in $M$
        }
        \ForAll{$m \in M$}{ 
            Create a new rule $r$ \\
            $r.h.where \gets  t.src$ \\
            \uIf(){$m \centernot= dummy$} {
            $r.h.when \gets  m$ \\
            }
            $r.body \gets  genRuleBody(r,dec\_p)$ \\
         Add $r$ into $R$ 
    }
    }
    }
    }
    \Return $R$  \\

\Fn{genRuleBody (Rule r, Decision point dec\_p)} {

    $O\gets \{t: t.src=dec\_p\}$ \\
     \uIf(){$r.h.where \in P6$} {
        $T_{tmp}\gets\{t: org(t).src=r.where\}$ \\
        $S_{tmp}\gets\{s: s = t.des \wedge t \in T_{tmp}\}$ \\
        $O\gets O \setminus  \{ t: t.src=dec\_p \wedge t.des \in S_{tmp} \}$ \\
     }
      
    \uElseIf(){$r.h.when \in P5$} {
       $O_{tmp}\gets O \cap  \{t: org(t)  \in P10 \wedge org(t).des \in P4 \}$ \\
       \uIf{$O_{tmp} = \emptyset$}{
            $O_{tmp} \gets O \cap \{t:   org(t)  \in P10 \}$ \\
        }\uIf{$ T_{tmp} = \emptyset$}{
            $O_{tmp} \gets O \cap \{t:  org(t).src \in P4 \}$ \\ 
        }
       
    }

    \uElseIf(){$r.h.where \in P3$} {
          $O_{tmp} \gets O \cap \{t: org(t).src \in P4 \}$ \\ 
     }
  $O\gets O_{tmp}$  \\
  \uIf(){$O = \emptyset$}{
    $O\gets \{t: t.src=dec\_p\}$ \\
  }
  Generate body of $r$ according to $O$
}

}
\caption{Generation of execution rules with a default body}
\label{ruleGeneration}
\end{algorithm}

\subsubsection{Generation of Default Execution Rules}
As discussed, at each decision point that has been added during the refinement (see Section~\ref{sec:refinment}), users are given a set of all possible options, one of which needs to be selected to continue the execution of the model. Considering and selecting one of the options can be time-consuming, specifically for the execution of large partial models. To minimize the efforts for making decisions, we generate default execution rules for the decision points that filter out options that are less likely to be selected by users. When the filtering results in a single option, the generated rule executes the model without user intervention, otherwise one of the remaining options needs to be selected by the user either interactively or by editing the generated rule. Note that the default execution rules are generated as a script and can be viewed, modified, or even ignored by users depending on the execution scenarios.

Algorithm~\ref{ruleGeneration} presents the method for the generation of default execution rules. The algorithm accepts as input the refined model and the partial elements which have been detected by the static analysis (\textit{P1-P11} as discussed in Section~\ref{sec:staticanalysis}) and then it generates a set of execution rules. 

For each decision point $dec\_p$, the algorithm iterates over all transitions that end at $dec\_p$ (line\# 4-5) and creates at least one rule for each transition (line\# 4-12). The \textit{where} part of each rule is set to the state the transition originated from, and the \textit{when} part is set to a guard indicating the arrival of a message $m$ triggering the transition (if any). 
Note that no rule is created for message \textit{$dbg$} (i.e., the debugging message). Thus, when the default rules are used, the execution can still be steered to any specific state by injecting the debugging message. Finally, the algorithm calls the function \textit{genRuleBody} (line\# 13), which generates a body for the rule.

Let us assume that when applied to a transition $t^\prime$ that may have been refined by Algorithm~\ref{superHSM}, the function $org(t^\prime)$ returns the original transition $t$ before the refinement. 
Function \textit{genRuleBody} applies the following three heuristics to filter out options less likely to be of interest to the user and generates the body of the rule. 

\begin{enumerate}

    
    
    \item If the rule handles a choice-point (i.e., the \textit{where} part of the rule is a choice-point such as $ch\_1$) with non-exhaustive guards, then the algorithm omits the transitions into states that are reachable via the transitions leaving $ch\_1$ (line\# 20-23) in the original model. The rationale is that the user has already decided when those states are to be reached from $ch\_1$ by specifying the guards of the outgoing transitions from $ch\_1$ in the original model.

    \item If the rule handles an unexpected message (i.e., the guard in the \textit{when} part of the rule indicates the arrival of an unexpected message) (line\# 24-29), the algorithm performs one of the following actions.
    (a) It selects not-takeable transitions that end at an isolated state, if any (line \#25); (b) otherwise, it selects not-takeable transitions, if any (line \#26-27); (c) otherwise, it selects transitions that end at an isolated state (line \#28-29); and (d) otherwise, it selects all possible transitions (line \#33-34). 
    
    Note that selecting a transition for an unexpected message has the same effect as adding the unexpected message as the trigger to the transition. The rationale for this heuristic is that the trigger of not-takeable transitions is missing, and setting the trigger for them is likely more helpful to make the model executable than the appending of a new message to the trigger of transitions that are already defined in the original model. 
    Similarly, since an isolated state does not have any incoming transition, there is more of a need to fix it compared to states that already have incoming transitions. 
    \item If the rule handles a broken chain, the algorithm tries items c) and d), in the same way as the previous heuristic (line\# 30-31). 
\end{enumerate}

We note that the generated rules may require further manual modification, due 
to the fact that a rule contains more than one option for 
selection, or the user finds one of the heuristics unsuitable
for their needs. However, the generated rules should provide a useful initial version even in these cases. 

For illustration, Listing~\ref{defaultexecurules} shows the default execution rules that are generated for the refined \textit{CTRSM} (see Fig.~\ref{fig:refinedctr}). Rules~\textit{r1-r2} are examples for the second heuristic, and rules~\textit{r3-6} filter out not-takable transitions $t_{13}$ by explicitly selecting transition $t_{r1}$. Note that rules~\textit{r1-r2} filter out all options except one, and therefore users may not need to change them or provide inputs at runtime interactively. However, rules~\textit{r3-r6} still select more than one option, and therefore the user still needs to update them or provide interactive input at runtime.


\begin{algorithm*}
\SetKwInOut{Input}{Input}
\SetKwInOut{Output}{Output}
\SetAlFnt{\scriptsize\sf}
\SetKwProg{Fn}{Function}{}{}
\DontPrintSemicolon
{

\Fn{saveDecesionsAsRules (A sequence of execution record L)}{
Check and resolve the inconsistencies of the decisions in $D$ \\
Let $R$ be an empty set \\
\ForAll(\tcp*[h]{l is an execution record (ref. Def.~\ref{def:execrec})}){$l \in L $}{ 
    $r \gets \{r \in R : r.h.where=l.c.\gamma.\sigma \wedge r.h.when=receipt(l.c.m) \}$  \tcp*[h]{l.c.p refers to a problematic element where the execution is recorded. } \\
    \uIf(){$r = \emptyset$}{
        create a rule $r$ \\
       $r.h.where=l.c.\gamma.\sigma$ \\
       $r.h.when= receipt(l.c.m)$ \\
        Add $r$ into $R$
    }
    Set body of $r$ based on the recorded debugging commands ($l.d$)  \\
}
 \Return $R$
}

\Fn{saveRuleToModel (\textit{HSM}\ sm, \textit{execution rule}\ $r$)} {
    
    Let $P1-P11$ refer to the problematic elements of $sm$ \\
    Let $selTrans$ and $selStates$  be the selected states and transitions by the rule extracted from the  $r.body$ \\
    \uIf(){$selStates$ has one member $\wedge$ $r.h.where$ refers to a state in $sm$} {
    Let $des$ be a state equal to the only member of $selStates$ in $sm$ and $src$ be a state equal to $r.h.where$ in $sm$ \\
    Let $selTran$ be the first member of $selTrans$ \tcp*[h]{ $selTrans$ has maximum one member since  $selStates$ has one member and for a state, more than one transition can not be selected} \\
    \uIf(){$selTran \centernot \in P10$}  {  
    $t \gets add\_trans(src, des)$ \\
    }\uElse{
       $t \gets  selTran$ 
     }
    
    $t.act \gets $ ($r.body$ without $select$ statement) + $t.act$ \\
    Set $t.trig$ based on the $receipt$ statements from $r.h.when$ \\
    Set $t.guard$ based on the $r.h.when$ by excluding the $receipt$ statements \\
     \uIf(){$r.h.where \in P6$} {
        $t.guard \gets   t.guard \wedge \big(\neg \bigvee_{t'\in out\_trans(t.src)} guard(t')\big)$  \\
     }

        
    }
}

}
\caption{Saving user decisions as execution rules and application of execution rules into design model}\label{savedecesions}
\end{algorithm*}

\begin{figure}
\begin{lstlisting}[caption={Default execution rules that are generated for the refined \textit{CTRSM} (see Fig.~\ref{fig:refinedctr})},captionpos=b,label=defaultexecurules,language=execrule]
rule r1 where state off when  receipt(timeout) {
    select state off  using t13
    }
rule r2 where state off when receipt(off)   {
    select state  off using t13 
    }
rule r3 where state yellow receipt(timeout) {
    select state  red|green|yellow|off
    }
rule r4 where state yellow receipt(off) {
    select state  red|green|yellow|off
    }
rule r5 where state red receipt(off) {
    select state  red|green|yellow|off
    }
rule r6 where state green receipt(off) {
    select state  red|green|yellow|off
    }

\end{lstlisting}
\end{figure} 

\subsubsection{Save user decisions (inputs) as execution rules}
As discussed, via interactive execution, users need to provide input to steer the execution at decision points. Depending on the goals of the execution, users may need to repeat the execution, and therefore providing the same input can be time-consuming and tedious. To deal with this issue, 
 a feature is provided that allows users to save interactive decisions in the form of execution rules. Thus, the execution can be repeated based on the saved rules without having to provide the input again. The implementation of the feature is heavily dependent on appropriate support for recording and viewing the execution of a partial model together with the user decision (input) provided during the execution. In the following, we discuss a high-level overview of how this feature is realized. A central notion is that of an execution record.

\begin{definition} \textit{(Execution Records)} \label{def:execrec} 
Let us assume that the execution of a partial model is saved as a set of records $\langle c,d, o  \rangle$, where $c$ denotes the execution context at the time of the decision,  $d$ refers to a sequence of debugging commands that have been issued by the user while the execution was stopped (i.e., from when the decision point was first reached and until the execution is resumed with a \textit{select} command), and \textit{o} refers to the runtime decision (input) that is taken by the user (i.e., the argument of the \textit{select} command). 
\end{definition}

Function \textit{saveDecesionsAsRules} in Algorithm~\ref{savedecesions} presents how the interactive decisions are saved as execution rules. It accepts a set of execution records and returns a set of execution rules. The function first checks that the decisions are consistent (i.e., unique decisions are taken for the same execution contexts) and then resolves inconsistencies by consulting with users, as will be discussed below. Then, it finds the rule that matches the execution context, and if no rule is matched, it creates an execution rule based on the execution context. Finally, it generates the body for the execution rule by using the issued debugging commands that modify the execution state (e.g., changing a variable value). 
Note that the interactive debugging commands (i.e., 'dbgCommands' and 'umlrtCmd' in Listing~\ref{dbgservices}) are a subset of the statements used for writing the body of an execution rule and there is no mismatch that complicates the use of sequences of debugging commands as rule bodies.

\textbf{Resolving the inconsistency.} Users are allowed to view their previous decisions and save one or more decisions as rules. When a user wants to save more than one decision, it is possible that some of the decisions are not consistent with each other, such that different decisions are taken at the same decision point for the same execution context. Thus, saving these decisions can cause non-determinism and should be avoided. To resolve inconsistencies between decisions, we ask users to select one of them. 


\subsubsection{Application of execution rules into design model}
To mitigate the issue of \textit{double effort}, we allow users to apply the execution rules into the design model automatically. 
To do that, we use the information in the execution rules about how to resolve partialness at runtime to fix the partialness in the design model. This feature is useful when users are satisfied that the way to deal with partialness at runtime expressed in the rules is correct and final, and they want to fix and remove partialness in the design model. 

Application of execution rule into the design model is addressed by function \textit{saveRuleToModel} in Algorithm~\ref{savedecesions}. It accepts the original state machine ($sm$) and an execution rule ($r$) as input and fixes the relevant partialness in $sm$ according to the definition of $r$, when there is only one possible solution to fix the partialness that is addressed by $r$ and $r.where$ explicitly refers to a state, i.e., $r.where$ does not contain $*$ or does not refer to a component. The function first extracts the selected states and transitions based on the $select$ statements from the rule's body, e.g., the selected states of rule $r1$ and $r3$  in Listing~\ref{execrulescriptsample} are $\{\textit{off}\}$ and  $\{\textit{red}, \textit{green}, \textit{yellow}, \textit{off}\}$ and the selected transitions are $\{t13\}$ and $\{\}$ respectively. It then takes the following steps. 
\begin{enumerate}
    \item { It checks if the rule selects only one state. 
    This check is necessary, because the application of a rule that selects more than one state can make the resulting model non-deterministic. 
    E.g., rule \textit{r6} in Listing~\ref{execrulescriptsample} selects more than one state (\textit{red, green, yellow, off}), and therefore saving this rule into the design model would require adding four transitions from state \textit{green} into the mentioned states with the same trigger (\textit{off}) and would cause non-determinism.} The function also checks if the \textit{where} part of the rule explicitly refers to a state because otherwise, it is not clear which element in the design model should be fixed (line\# 16). 
    \item It creates a transition ($t$) whose source is set to the $where$ part of the rule and whose destination is set to the state that is selected by the rule if the rule does not address a not-takeable transition ($P10$ partialness). There is no need to add a new transition to fix a not-takeable transition (line\# 17-22). Note that if the source and destination states of $t$ are not contained in the same composite state, function $add\_trans$ adds required transitions and related entry-point and exit-point states to assure the transition does not cross the boundary of its parent state (see Def.~\ref{def:weelformed}).  
    \item It then sets the action of $t$ based on the body of the rule. Note that since a not-takeable transition may already have actions. Therefore the action is appended to the body of the rule. The trigger of $t$ is set based on the $receipt$ statements in the \textit{when} part of rules, and its guard is set based on the \textit{when} part of the rule, excluding the $receipt$ statements (line\# 23-25).
    \item  Finally, if the rule handles the non-exhaustive guard of a choice point $ch_1$, the guard of $t$ is conjoined with the negation of the disjunction of the guards of all transitions leaving $ch_1$, as calculated during the refinement (line\# 26-27). 
\end{enumerate}

{
Note that when a rule is applied, it fixes the relevant partialness and there is no need to manually modify the model after its application, because the model is updated in a way that follows the semantics of the rules precisely, and respects all \textit{HSM} well-formedness constraints (see Def.~\ref{def:weelformed}). Also, the algorithm only presents the application of an execution rule into the design model. Saving an interactive decision into the design model can be performed by saving it as an execution rule, which is then can be applied to the design model as discussed. 
}

\subsection{Tool Support (PMExec)}
\label{sec:PMExec}

We have developed \textit{PMExec}\footnotemark\footnotetext{\url{https://moji1@bitbucket.org/moji1/partialmodels.git}} that embodies our approach and supports execution of partial UML-RT models. We used the Epsilon Object Language (EOL)~\cite{epsilon2008} to implement the transformation rules required for refining the models into executable models. EOL supports a set of instructions to create, query, and modify models. The part for the execution of the refined models (debugging probe) is implemented using C++, ANTLR~\cite{antlr}, and the Boost C++ Library ~\cite{boost}.


\subsubsection{PMExec Features}
\begin {figure*}[t!]
\centering
\begin{tikzpicture}

\node[anchor=south west,inner sep=0] (image) at (0,0) {\includegraphics[width=0.80\textwidth]{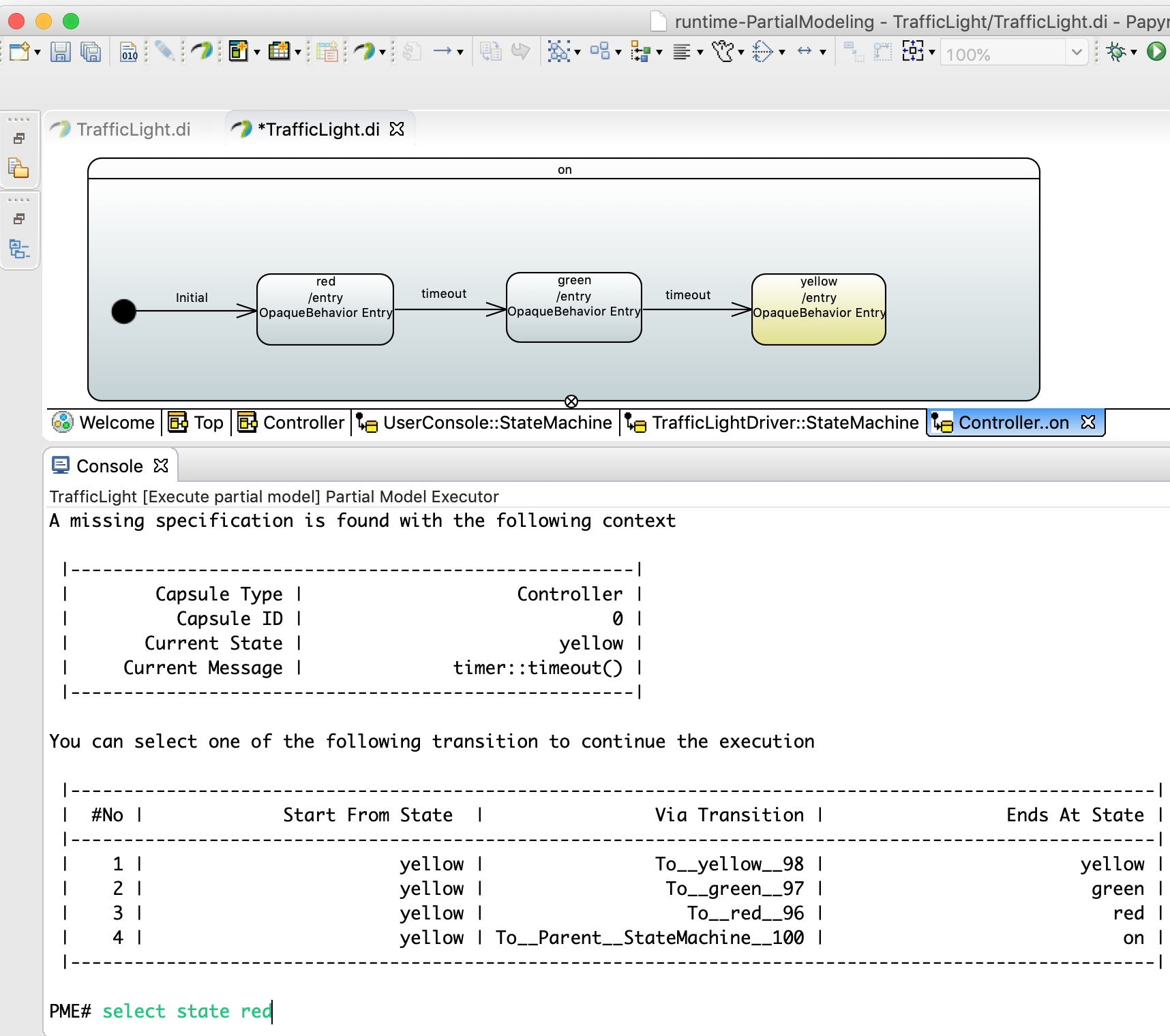}};
\begin{scope}[x={(image.south east)},y={(image.north west)}]
\node[anchor=west,circle,drop shadow, fill=blue, text=white, inner sep=1pt, font=\normalsize] (mark1) at (.5,.80) {1};
\node[anchor=west,circle,drop shadow, fill=blue, text=white, inner sep=1pt, font=\normalsize] (mark1) at (.5,.50) {2};
\end{scope}
\end{tikzpicture}%
\caption{User interface of PMExec}
\label{fig:GUI}
\end{figure*}

In the following, we discuss the features of \textit{PMExec}\footnote{A demonstration video can be found at \url{ https://youtu.be/BRKsselcMnc}} from the user point of view. When it is possible, the use of features is explained using the running example.

\noindent\textbf{Setup and run} The PMExec is integrated into Papyrus-RT as an Eclipse plug-in and can be downloaded and installed from the \textit{PMExec} repository. After installation, it can be used to run partial UML-RT models simply by defining a run configuration (i.e., an Eclipse run configuration) inside Papyrus-RT. The static analysis, transformation, code generation and build run automatically in the background without distracting the user. Upon  successful execution, \textit{PMExec} loads a UI as shown in  Figure~\ref{fig:GUI} as soon as the execution requires user input to continue the execution. The UI is split in two parts, a \textit{HSM view} (\token{1}{blue} of Figure~\ref{fig:GUI}) and a \textit{DBG} console (\token{2}{blue} of Figure~\ref{fig:GUI}). In the HSM view the user can see the HSM of the capsule where the current execution state is highlighted. In the \textit{DBG} console the user can interactively issue commands to investigate and fix the execution problems. Some of the most important commands are discussed in the following in the context of the running example. 

\textbf{View/select options} List the possible options to fix/continue the execution. E.g., the output of \textit{view options} for the \textit{CTR} when its execution is stuck in state \textit{yellow} is shown in part \token{2}{blue} of Figure~\ref{fig:GUI}. Using the console, the execution can now be steered to any of the defined states inside the HSM. The command \textit{select} allows users to select one of the possible options, e.g., `\lstinline|select state red|' steers the execution to state \textit{red}.

\textbf{Simple expressions} Similar to scripting languages (e.g., Python's interactive console), \textit{PMExec} allows the user to issue simple expressions (e.g., arithmetic expressions) and statements (to, e.g., define a new variable, or change/view variable values). This allows the user to investigate and modify the execution before deciding how to advance the execution. E.g., `\lstinline|x=5+1|' creates a new variable \lstinline|x| and sets its value to $6$. Defined variables can help the user record certain properties of the execution and define complex debugging and testing scenarios. Once defined, they can be used till the end of the execution. 

\textbf{Communication commands} To allow the user to fix issues concerning the missing inputs (\textit{P7}), three communication commands are provided: \textit{inject}, \textit{send}, and \textit{reply}. The command \textit{inject} sends a signal to a capsule to start a debugging session, the command {send} sends messages on behalf of the capsule being debugged to the connected capsules, and the command \textit{reply} sends an incoming message back on the same channel it has been received. E.g., in the context of running example, no behavior is defined for the component \textit{UC}. Thus, the execution of the \textit{CTR} will get stuck in state \textit{off} and the overall execution of the system will be deadlocked. The user can fix the problem by using the following communication command (1) `\lstinline|inject UC|' to start a debugging session with capsule \textit{UC}. Note that the refinement fixes the behavior of capsules even with no defined behavior. (2)  `\lstinline|send message on|' to send  message \textit{on} to the \textit{CTR} where it will trigger a transition to turn on the \textit{red} light.

\begin{figure}
\begin{lstlisting}[caption={An example script of the execution rules in the context of the \textit{TrafficLight}},captionpos=b,label=execrulescriptsample]
rule r1 where state yellow when  receipt(timeout) {
    select state  red}
rule r2 where component * {
    reply  random
    select state random }
\end{lstlisting}
\end{figure}

 \textbf{Batch execution}   \textit{PMExec} supports a batch execution mode which allows users to provide inputs using a script of execution rules. The Listing~\ref{execrulescriptsample} is an example of an execution script in the context of the running example. 
 \begin{enumerate}
     \item The rule \textit{r1} steers the execution to state \textit{red} when a message \textit{timeout} is received while in state \textit{yellow}. 
     \item The rule \textit{r2} replies to any received message using a random message and then moves the execution to a random state. The rules with header `*' are only selected when no other rule matches in the current execution context. Note that having only one rule similar to \textit{r2} is enough for the random execution of any partial model using \textit{PMExec}.
 \end{enumerate}
 
\lstset{language=C}

  \textbf{Save command} To save users' decisions which are provided interactively as execution rules, users can view the history of the execution using `\lstinline|view exec|'. The output shows all previous decisions (inputs), each of which is given a unique id. Then, the user can use \textit{save} command (`\lstinline|save input id |') to save the input with the related id as an execution rule. Also, to save an execution rule into the design model, users can use `\lstinline|save rule id |' that saves the rule with the related id into the design model. Note that in both cases (\textit{save input/rule}), more than one input/rule can be processed by providing more than one id. 


\section{Validation}
\begin{figure}
    \centering
    \includegraphics[width=4.6cm]{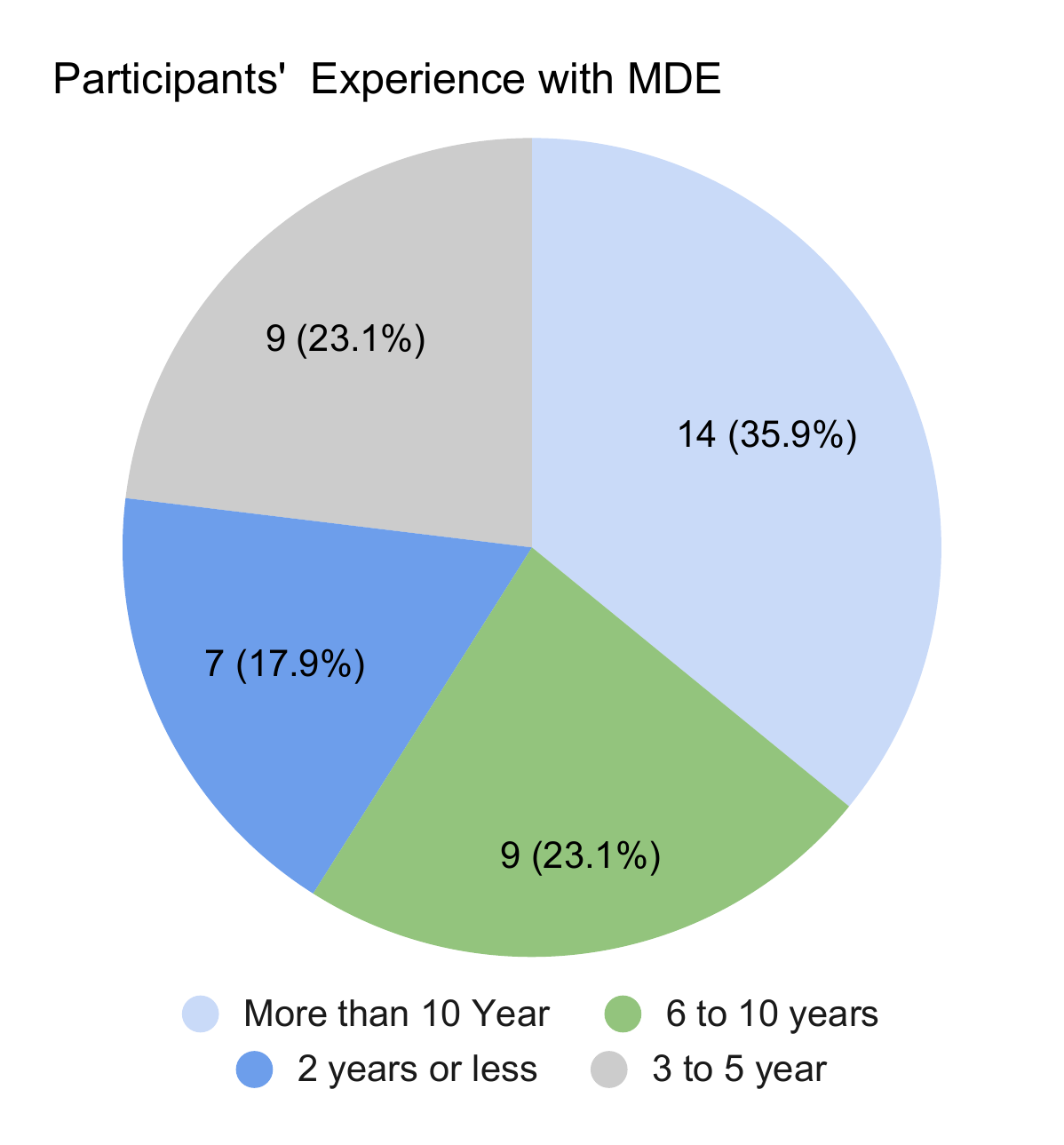}
    \includegraphics[width=3.95cm]{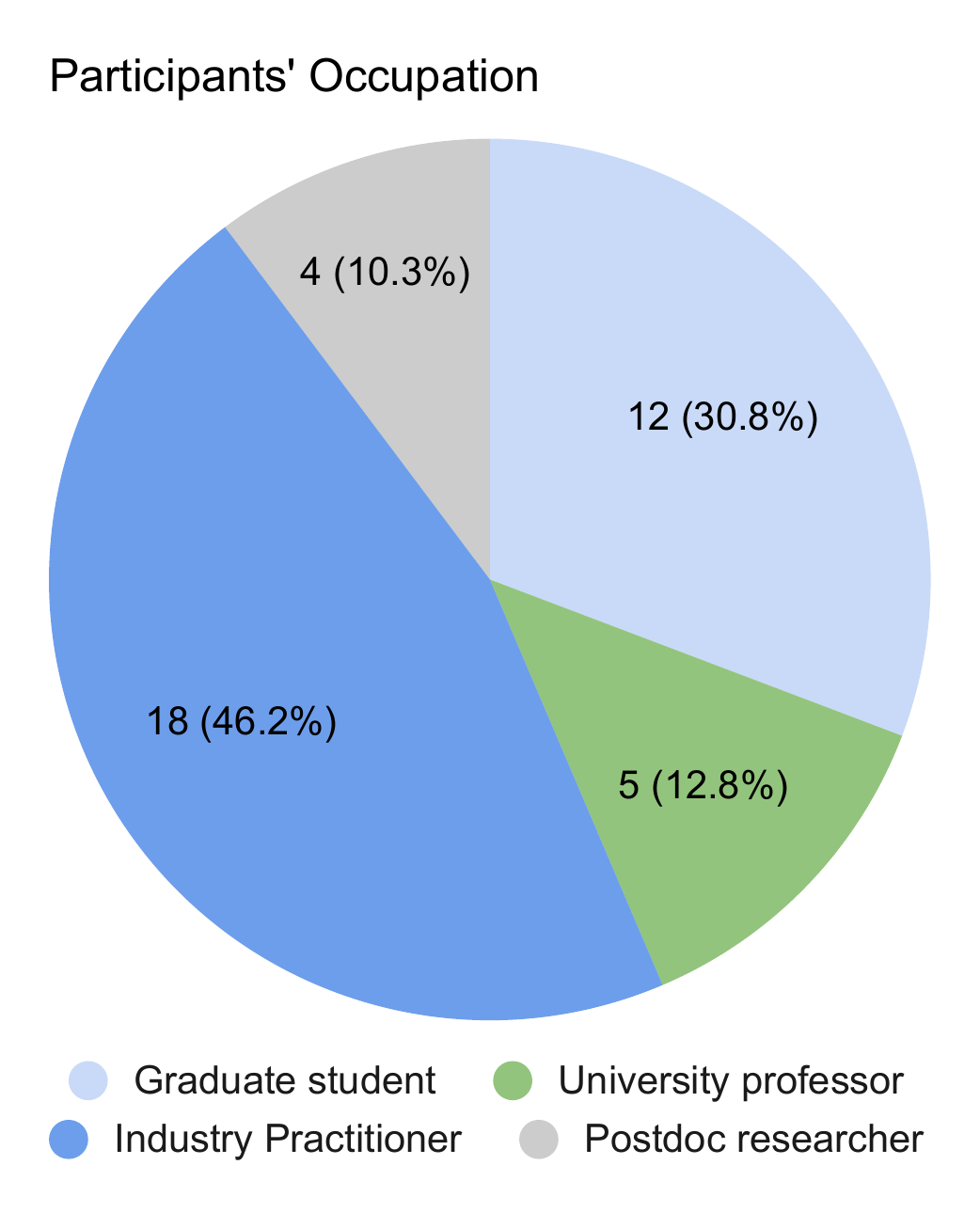}
    \caption{Survey participants'  occupation and experience with MDE}
    \label{fig:surveyparticipant}
\end{figure}

\begin{figure*}
    \centering
    \includegraphics[width=18cm]{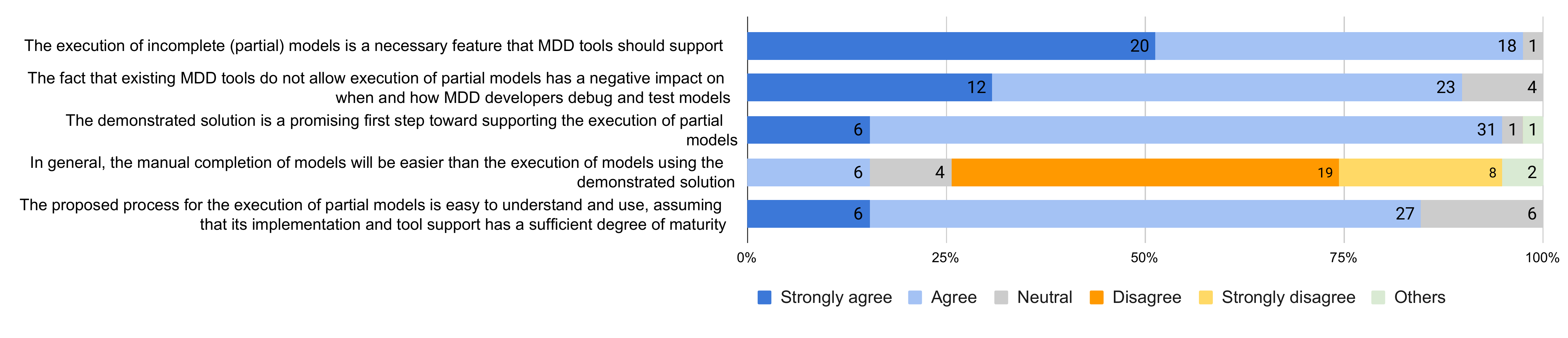}
    \caption{Participants' opinion concerning the execution of partial models and our proposed solution}
    \label{fig:surveyresults}
\end{figure*}

\begin{figure}
    \centering
    \includegraphics[width=9cm]{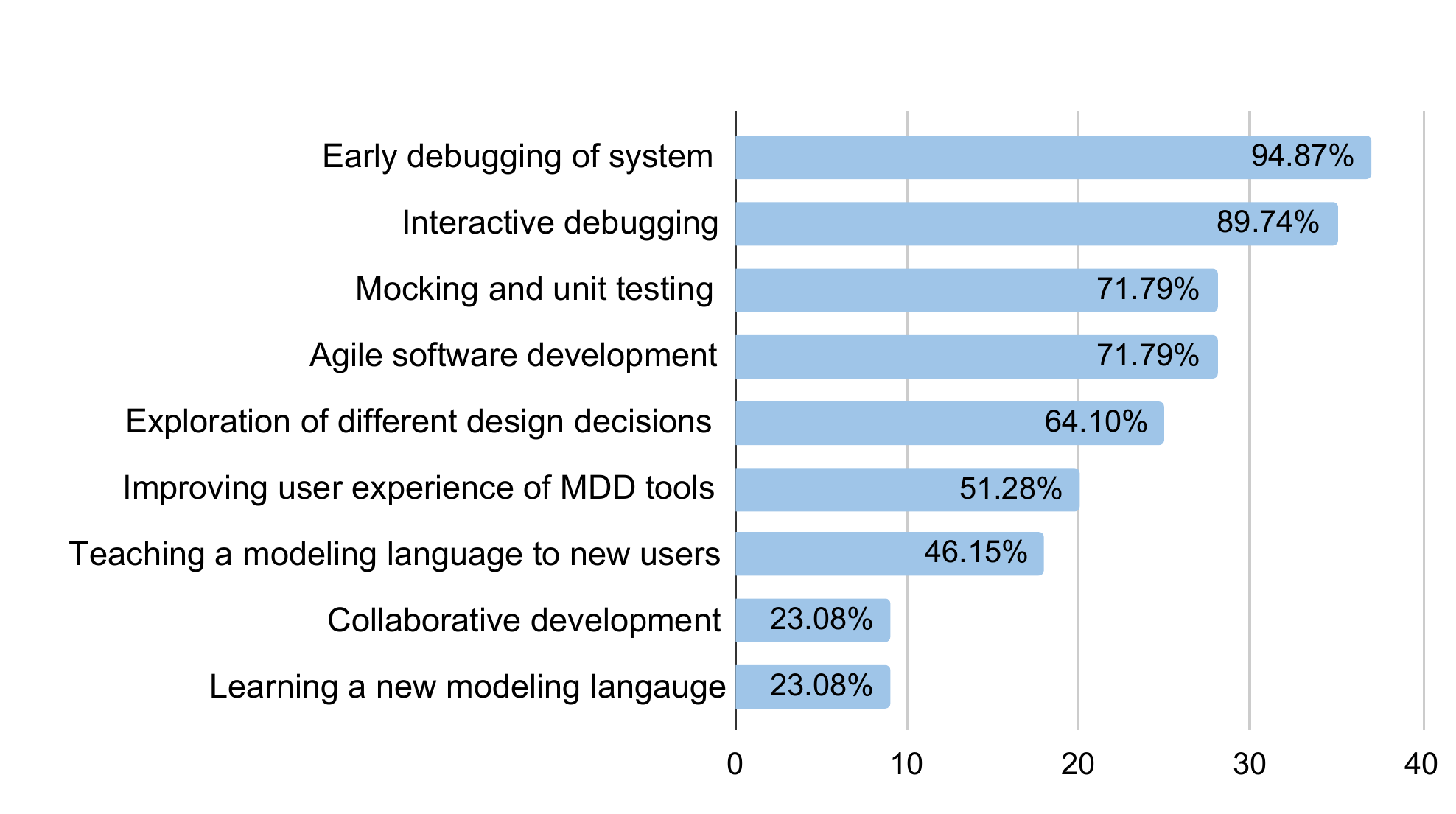}
    \caption{Participants' opinion about which activities the execution of partial models can facilitate}
    \label{fig:usefullnessofPME}
\end{figure}

\label{sec:evaluation}
This section explains the validation of our approach which consists of three parts: \textit{online survey}, \codee{formal validation} and \codee{empirical evaluation}. The goal of the \textit{online survey} was to collect the opinions of MDE researchers and practitioners w.r.t.\ whether or not (1) the execution of partial models is a necessary and useful technology in the context of MDE, and (2) our approach is helpful to address the execution of partial models. The \textit{formal validation} is concerned with the properties of the refinement approach and shows formally how the applied refinement does not change the behaviour of the original specification of the models but fixes the problems of lack of reachability and progress. The \codee{empirical evaluation} is concerned with the applicability of the approach in practice. It applies our approach to several partial UML-RT models in different scenarios and evaluates the performance and overhead. In the following, we discuss each part in detail.

\subsection{Online survey}

\subsubsection{Survey Design} The survey\footnote{\url{https://tinyurl.com/yxv8embf}} consists of two steps: First, we ask participants who are MDE researchers or practitioners to view a short demonstration video (5 minutes) of \textit{PMExec} to familiarize them with the execution of partial models. { Note that the video does not demonstrate the automation features as discussed in Sec.~\ref{sec:Automation}, since the features were inspired by the suggestions of the participants.} Second, we ask them to answer 15 questions classified into three groups: {demographic} ($4$ questions), general questions regarding the execution of partial models ($5$ questions), and specific questions concerning our proposed approach ($6$ questions). $7$ questions are 5-level Likert scale questions~\cite{albaum1997likert} (\textit{Strongly disagree}, \textit{Disagree}, \textit{Neutral}, \textit{Agree}, \textit{Strongly agree}), 6 questions are multiple choice questions, one question is asking for the email address of the participant (\textit{email} question), and one question is an open-ended question.  All of the questions except the open-ended and \textit{email} question are mandatory. However, participants are allowed to provide other answers rather than the choices or scales presented to them. {Note that, providing an email address is optional to allow participants to be anonymous in case they want to provide negative feedback.}

\subsubsection{Participants}
We approached MDE practitioners and researchers in person at the MODELS (2019) conference, as well as by email and social media and asked them to participate in the survey. Our efforts resulted in $39$ participants whose demographic data is shown in Figure~\ref{fig:surveyparticipant}. More than 81\% of participants have more than two years of experience with MDE, 46\% of them work in the industry, 13\% of them are university professors, 10\% are postdoctoral research fellows, and 31\% of them are graduate students.  Notably, all of the participants except one are currently dealing with MDE tools in their work, and around 36\% of participants have more than 10 years of experience and can arguably be said to be internationally known experts in the field. Finally, only 13 (33\%) of participants provided their email addresses, and 26 (67\%) of them participated anonymously.

 \subsubsection{Results}
\textbf{The relevance of the execution of partial models.}
We have asked four questions from the participant to understand whether they think the execution of partial models is essential and how it can help  developers. 
As shown in Fig.~\ref{fig:surveyresults}, except for one of the participants, all of them either strongly agree (51.3\%) or agree (46.2\%) that the execution of partial models is a necessary feature that needs to be supported by MDE tools. Also, as shown in Fig.~\ref{fig:usefullnessofPME}, participants think that the execution of partial models can be helpful for several software development activities, mainly early and interactive debugging, unit testing and agile software developments. Also, more than 51\% of participants think that the execution of partial models can improve the user experience of MDE tools.  Overall, since the vast majority of the participants perceive the execution of partial models as a relevant technology,  we can safely conclude that addressing of the execution of the partial models is a relevant problem and worth addressing.

\textbf{The usefulness of our approach} \label{sec:surveyresult}
As shown in Figure~\ref{fig:surveyresults}, all of the participants except two of them perceive our approach as a first promising step toward addressing the execution of partial models. Also, 84.4\% of the participants either strongly agree (15.4\%) or agree that the current approach can be useful for MDD users, assuming that the approach has good tool support.

86.4\% of the participants prefer to use both interactive and scripting methods for providing input, depending on the execution scenarios.  69.2\% of participants are of the opinion that the execution of models by providing input through scripts is easier than through a manual completion of models. However, 15.4\% of participants have the opposite view.

{
In addition, we received valuable and constructive responses to the open-ended question as discussed in the following.  

1) As quoted in the following, one of the participants pointed out correctly that our approach only is applicable to modeling languages with step-based execution semantics. \\
\textit{“This can likely work well with behavioral models such as statecharts or business process models, but I wonder about whether the value proposition also covers non-behavioural models such as goal models, where "execution" is not a sequence of events but a set of values or initial decisions”}

2) As discussed, the video does not demonstrate the automation features as mentioned in Sec.~\ref{sec:Automation}, since the following suggestion inspired the features.

\textit{“The script option sounds like it does not much improve on manual fixing/completion of models. I suspect that it may prove more useful if: (a) the script is automatically generated (as a user option) during an interactive session and then applied (as a user option) on subsequent runs. (b) used as a means of automatically modifying the incomplete model --- again, as a user option.”}

3) As quoted in the following, one of the participants suggested an interesting extension to the work by augmenting the semantics of modeling language to support the execution in the presence of holes that are specified with certain notations. We agree with the participant. However, we left addressing this extension to future work. 

\textit{“The partiality of the model in the demo seems to pertain only to violated statically checkable constraints such as whether a state is exitable/reachable. This would certainly help modellers.
Another type of partiality is the presence of "holes", i.e. properties not being filled in (cf. Scala's "???"), not resolved, etc. I think it's a good idea in general to augment semantics of a modelling language that they stay defined (but possibly defined in terms of a fault mode) in the presence of missing model parts. This is akin to interpret every .-operator as ?.-operator (Kotlin, TS, etc.) and propagate nulls/undefined in a as meaningful as possible manner.”
}

4) Not surprisingly, as quoted in the following, two of the participants have constructive feedback concerning the tooling issues. However, in this work, our main focus is the creation of a prototype as proof of concept, and improving of tooling is left to future work.

\textit{“I found your presentation of the context to be rather technical. For example, at 2 minutes in the video you showed the output for the empty UserConsole. There you presented one option how I could continue. I think you might have a technical reason to call the states Init\_\_State\_\_3 and New\_\_State\_\_2 but to me as a user it is unclear what these numbers mean (any why a state that is not even created yet(?) would have a lower number than the initial state you already generated). But I really liked, that you highlighted the current state in the visual representation of the diagram.”}
}

Overall, based on the participants' opinions, we can conclude that while our proposed approach is a step in the right direction, it is not the final solution (i.e., it has limitations). Still, further research and development are required in this context, some of which will be discussed in Section~\ref{sec:discussion}.

\subsection{Formal Validation} \label{sec:proofs}

We use $\Re^{HSM}$ to refer to the result of applying Algorithm~\ref{superHSM} on an \codee{HSM} and call it the \codee{RefinedHSM} of the \codee{HSM}. In the following, first, we define the simulation relationship between \codee{LTSs}, and then discuss the properties of $\Re^{HSM}$.   \\

\subsubsection{Behavioural Preservation}
\begin{definition} (Simulation Relation) 
Let $L_1 = \langle \Gamma_1, \mathcal{A}_1, {\gamma_1}_{0}, \mathcal{Q}_1, \rightarrow_1 \rangle$  and $L_2 = \langle \Gamma_2, \mathcal{A}_2, {\gamma_2}_{0}, \mathcal{Q}_2, \rightarrow_2 \rangle$ refer to LTSs of two HSM, $HSM_1$ and $HSM_2$ respectively (LTSs are discussed in detail in Def.~\ref{sec:semantic}). We write $L_1 \preceq L_2$ and say  $L_2$ simulates $L_1$ if there is a binary  relation ${R} \in$  $\Gamma_1 \times \Gamma_2$ with the following two properties. \\
\noindent \textbf{Start property:} 
${\gamma_1}_{0} \centernot = \emptyset \implies$ $({\gamma_1}_{0}, {\gamma_2}_{0}) \in {R}$. Note that, in the execution of HSMs only one initial state is allowed. \\
\noindent \textbf{Step property:} Let  $(\gamma_1,\,\in \Gamma_1$,\, $\gamma_2 \in \Gamma_2) \in R$. 
\begin{align*}
\begin{tabular}{ll}
For all  $ \gamma_1^\prime \in \Gamma_1$  and $a_i,$ whenever  $\gamma_1 {\xrightarrow[]{a_1 \cdots a_n}}_1 \gamma_1^\prime$  \\ there exist $ \gamma_2^\prime \in \Gamma_2$ such that $\gamma_2 {\xrightarrow[]{a_1 \cdots a_n}}_2 \gamma_2^\prime \,$  $\wedge \, (\gamma_1^\prime, \gamma_2^\prime) \in {R}$. & 
\end{tabular}
\end{align*}

 The step property implies that when $(\gamma_1, \gamma_2) \in {R}$, any execution step started from  $\gamma_1$ can be matched by an execution step started from $\gamma_2$ such that they both execute the same actions and reach configurations that again are in relation $R$. 

Simulation implies trace containment, i.e., every sequence of actions that is possible by the simulated $LTS$, is also possible by the simulating $LTS$~\cite{Lynch:1996:DA:2821576} (i.e., the simulating $LTS$ preserves the specification of the simulated $LTS$).   
\end{definition}

\begin{definition}
\label{def:R}
Let LTSs $L_o=\langle \Gamma_o,\, \mathcal{A}_o,\, {\gamma_o}_{0},\, \mathcal{Q}_o,\, \rightarrow_o\, \rangle$ and $L_r=\langle \Gamma_r,\, \mathcal{A}_r,$ $\, {\gamma_r}_{0},\, \mathcal{Q}_r, \,\rightarrow_r \rangle$ represent the execution semantics of \codee{HSM} and $\Re^{HSM}$ respectively, $\Re^{{\mathcal{E}}}$ refers to a mapping from newly introduced variables during the refinement to their values, and  $\mathcal{R} \in \Gamma_o \times \Gamma_r$ is a binary relation defined as follows.
\begin{align*}
\begin{tabular}{ll}
 $\mathcal{R} = \{(\gamma_o \in \Gamma_o ,\gamma_r \in \Gamma_r) \, | \,  \gamma_o.\sigma = \gamma_r.\sigma \wedge \gamma_o.{\mathcal{E}} =  \gamma_r.{\mathcal{E}} \setminus \Re^{{\mathcal{E}}}$ \\  $ \wedge\, \gamma_o.\overline{\mathcal{H}} = \gamma_r.\overline{\mathcal{H}}\}$&  \\
\end{tabular}
\end{align*}

\end{definition}

\begin{prop}
\label{prop:simulation}
Assuming that $L_o$ and $L_r$ receive the same sequence of messages ($Q_r=Q_o$) and users do not issue any debugging commands during the execution of $\Re^{HSM}$, the relation $\mathcal{R}$ (as defined above) is a simulation relation, i.e., execution of an $\Re^{HSM}$ simulates the execution of \codee{HSM} ($L_o \preceq L_r$). 

\begin{lemma}
\label{lemma:nextt}
For message $\mu$ and basic state $s$, if function $next\_t(s,\mu))$ (see Table~\ref{tab:helperfunction}) returns transition $t$ in the context of \codee{HSM}, then it returns the same transition ($t$) in the context of $\Re^{HSM}$. 
\begin{proof} (Lemma \ref{lemma:nextt})
According to Algorithm~\ref{superHSM}, the refinement applies the following changes to the basic states: (\RNum{1}) Add a transition from a basic state to $dec\_p$. The trigger of this transition is set to $in(c) \setminus handled(s_{b})$  in order to not affect the existing transitions. This ensures that if a transition of \codee{HSM} can be triggered by message $\mu$, it still can be triggered by the same message in $\Re^{HSM}$ and $next\_t$ in both cases returns the same transition. (\RNum{2}) The source of not-takeable transitions is changed to $dec\_p$. $next\_t$ never returns a not-takeable transition, thus this change does not affect function $next\_t$. (\RNum{3}) A transition is added from $dec\_p$ to isolated states. Function $next\_t$ never returns an incoming transition to a state as the result. Thus, this change does not affect $next\_t$ either.
Based on (\RNum{1}), (\RNum{2}), and (\RNum{3}) the proof of this lemma is complete. 
\end{proof}

\end{lemma}

\begin{lemma}
\label{lemma:deadS}
For basic state $s$, if function $dead(s))$ (see Table~\ref{tab:helperfunction}) returns \codee{false} in the context of \codee{HSM}, then it also returns \codee{false} for state $s$ in the context of $\Re^{HSM}$. 

\begin{proof} (Lemma \ref{lemma:deadS})
No state or takeable transition is removed by the refinement. Thus if state $s$ or one of its ancestors (\codee{parents(s)}) has a takeable transition ($t$) that prevents $s$ from being dead, the same transition also exists in $\Re^{HSM}$. This completes the proof of this lemma.
\end{proof}
\end{lemma}

\begin{proof}(Proposition \ref{prop:simulation}) To prove that $\mathcal{R}$ is a simulation relationship, first, we need to show that the start property holds which includes the following two cases. 
\begin{itemize}
    \item The initial state of \codee{HSM} is missing. This case is trivial, since without initial state, the execution of \codee{HSM} cannot start (see Def.~\ref{sec:semantic}) and ${\gamma_o}_{0}=\emptyset$. Thus, the start property holds for this case.
    \item \codee{HSM} contains the initial state (i.e., ${\gamma_o}_{0}\centernot=\emptyset$). In this case we need show that $({\gamma_o}_0,{\gamma_r}_0) \in \mathcal{R}$ where ${\gamma_r}_0$ is the initial configuration of $\Re^{HSM}$.  (\RNum{1}) According to (lines~\# 5-6 of Algorithm~\ref{superHSM}), when the original \codee{HSM} contains the initial state $in_0$, the refinement keeps the same initial state in the refined HSM (i.e., the initial states of $\Re^{HSM}$ and \codee{HSM} are equal to $in_0$). According to execution semantics of \codee{HSM} (see Def.~\ref{sec:semantic}), the execution of \codee{HSM} starts from the initial configuration where its current state is set to the initial state of the \codee{HSM}. Thus, there is an initial configuration of ${\gamma_r}_{0} \in \Gamma_r$, in which the current state is equal to $in_0$  i.e., (${\gamma_o}_{0}.\sigma={\gamma_r}_{0}.\sigma$).  (\RNum{2}) ${\gamma_o}_{0}.\mathcal{H} = {\gamma_r}_{0}.\mathcal{H}$ because the history is set to empty for the initial configuration (see Def.~\ref{sec:semantic}). (\RNum{3}) Similarly, ${\gamma_o}_{0}.{\mathcal{E}} = {\gamma_r}_{0}.{\mathcal{E}} \setminus \Re^{{\mathcal{E}}}$, because the initial values of variables are set to default values and the refinement does not remove any existing variable. Based on (\RNum{1}), (\RNum{2}), and (\RNum{3}) we can conclude that (${\gamma_o}_0,{\gamma_r}_0) \in \mathcal{R}$ and conclude that the start property of simulation holds for this case as well.  
\end{itemize}

\noindent Second, we have to show that the step property holds for any (${\gamma_o} \in \Gamma_o, {\gamma_r} \in \Gamma_r)$ $ \in \mathcal{R}$ which includes two main cases according to the execution semantics of \codee{HSM} (see Def.~\ref{sec:semantic}).
\begin{itemize}
    \item $\gamma_o$ is a stuck configuration, i.e., no execution step can originate from it. Thus, the step property holds for this case. 
    \item $\gamma_o$ is a not a stuck configuration. Based on the execution rules (see Def.~\ref{sec:semantic}), this case includes 4 sub-cases: (1) the current state ($\gamma_0.\sigma$) is a pseudo-state of kind initial, entry-point, exit-point, or junction-point (Rule-1), (2) the current state is a basic state (Rule-4), (3) the current state is a history state (Rule-6),  and (4) the current state is a choice-point (Rule-8). Proof of all sub-cases is similar and here we only prove sub-case (2). \\
    Let us assume that ${\gamma_o}.\sigma \in \mathcal{S}_b $,  (${\gamma_o} \in \Gamma_o, {\gamma_r} \in \Gamma_r)$ $\in \mathcal{R}$ and an execution step $st_o=(\gamma_o \rightarrow_o \gamma_o^\prime)$ is taken. First, we have to show that an execution step $st_r=(\gamma_r \rightarrow_r \gamma_r^\prime)$ can be started from {$\gamma_r$} that executes the same actions as $st_o$. To prove the existence of $st_r$, we have to show that the following condition holds (see Rule-4, Def.~\ref{sec:semantic}).
    \begin{align*}
        \begin{tabular}{ll}
        $ \mathcal{Q}_r \centernot = \emptyset \, \wedge \,  \gamma_r.\sigma \in \mathcal{S}_b \wedge \neg dead(\gamma_r.\sigma) \, \wedge \,$ \\ $\exists\, t \in \mathcal{T}\,| \,  t=next\_t(\gamma_r.\sigma,head(\mathcal{Q}_r))$  
    \end{tabular}
    \end{align*} 
    (\RNum{1}) Definition of $\mathcal{R}$ and assumption $\gamma_o.\sigma \in \mathcal{S}_b$ imply that $\gamma_r.\sigma \in \mathcal{S}_b$. 
    (\RNum{2})  $Q_o \centernot = \emptyset$ because the execution step $st_o$ is not possible with an empty queue (see Rule-4, Def.~\ref{sec:semantic}).   Also, $Q_o=Q_r$ based on the assumption of the proposition. Thus, $\mathcal{Q}_r \centernot = \emptyset.$ 
    (\RNum{3}) $\neg dead(\gamma_o.\sigma)$ holds, because without that execution step $st_o$ is not possible. Thus, based on Lemma~\ref{lemma:deadS} $\neg dead(\gamma_r.\sigma)$ holds.
    (\RNum{4}) First, $\exists\, t \in \mathcal{T}\,| \, t=next\_t(\gamma_o.\sigma,head(\mathcal{Q}_o))$ holds, otherwise the execution step $st_o$ is not possible. Second, ${Q}_r = Q_o$ implies that  $head({Q}_r)= head({Q}_o)$. Third, $\gamma_o.\sigma = \gamma_r.\sigma $. Based on the statements above and Lemma~\ref{lemma:nextt}, we can conclude that the last part of the above formula ($\exists\, t \in \mathcal{T}\,| \, t=next\_t(\gamma_r.\sigma,head(\mathcal{Q}_r))$) holds. Based on \RNum{1}, \RNum{2}, \RNum{3}, and \RNum{4} we conclude that execution step $st_r$ exists.
    
    Next, we have to show that $st_r$ and $st_o$ execute the same sequence of actions. 
    According to Rule-4 (see Def.~\ref{sec:semantic}), definition of relation $\mathcal{R}$, and Lemma~\ref{lemma:nextt}, both $st_r$ and  $st_o$ execute the same actions, i.e, $exit(t.src),exit(up\_s(s, t)),act(t),\, entry(t.des))$, where $s=\gamma_o.\sigma=\gamma_r.\sigma$, $\mu = head(\mathcal{Q}_o)=head(\mathcal{Q}_r)$ and $t=next\_t(s,\mu)$.
    
    Finally, we have to show that $(\gamma_o^\prime, \gamma_r^\prime) \in R$. (\RNum{1}) We have already shown that $next\_t$ returns the same result for both current states $\gamma_o.\sigma $ and $\gamma_r.\sigma$. Thus $\gamma_o^\prime.\sigma = \gamma_r^\prime.\sigma$ with $t.des$ as active state where $t$ is the result of $next\_t$ (Rule-4). 
    (\RNum{2}) Similarly, $\gamma_o^\prime.\mathcal{H}=\gamma_r^\prime.\mathcal{H}$ which are set to $u\_h(t.des,H)$. $\mathcal{H}$ refers to history of $\gamma_o$ and $\gamma_r$ which are equal (Def.~\ref{def:R}).
    
    (\RNum{3}) Variables are only changed by the execution of actions. Since the same sequence of actions is executed by $st_0$ and $st_r$, and $\gamma_o.{\mathcal{E}} = \gamma_r.{\mathcal{E}}\setminus \Re^{\mathcal{E}}$, we have $\gamma_o^\prime.{\mathcal{E}} = \gamma_r^\prime.{\mathcal{E}}\setminus \Re^{\mathcal{E}}$. Based on (\RNum{1}), (\RNum{2}), and (\RNum{3}) we can conclude that (${\gamma_o^\prime} \in \Gamma_o ,{\gamma_r^\prime} \in \Gamma_r)$ $\in \mathcal{R}$. Thus, the proof for sub-case (2) is complete.

\end{itemize}

    The other sub-cases can be proven similarly, and we conclude that the relation $\mathcal{R} \in \Gamma_o \times \Gamma_r$ is a simulation relation and by that execution of $\Re^{HSM}$ simulates the execution of \codee{HSM}.

\end{proof}

\end{prop}

While $\Re^{HSM}$ preserves the behavior of the original \codee{HSM}, it also never gets stuck and provides useful features to steer the execution to relevant states and debug the execution of the \codee{HSM}, assuming the required inputs are provided. In the rest of this section, we discuss the important properties of $\Re^{HSM}$.\\

\subsubsection{Reachability of the execution}

\begin{prop}(Reachability of States) 
\label{prob:reach1}
Assume $L_r=\langle \Gamma_r,\, \mathcal{A}_r,$ $\, {\gamma_r}_{0},\, \mathcal{Q}_r, \,\rightarrow_r \rangle$ is the execution semantics of $\Re_{HSM}$.
Let $\gamma$ be the current configuration of $L_r$, where $\gamma.\sigma \in \mathcal{S}_b$. By injecting a $dbg$ message, the execution can be steered by a finite number of execution steps to any configuration $\gamma^\prime$ in which the current state is any state except initial, choice-points, and composite (implicit history) states, assuming that proper inputs are provided.
\begin{proof} (Reachability of States)
Let $\sigma$ be the current state of configuration $\gamma$, and $\sigma^\prime$ be the current state of configuration $\gamma^\prime$ to which we want to steer the execution. Proving that $\gamma^\prime$  is reachable by taking a finite number of execution steps, includes three cases. (\RNum{1}) Both states $\sigma$ and $\sigma^\prime$ have the same parent, i.e., $parent(\sigma)=parent(\sigma^\prime)$. In this case, based on the execution semantics of $HSMs$, injecting a $dbg$ message starts an execution step that moves the execution to $dec\_p$ with the same parent (line\# 12-14 of Algorithm \ref{superHSM}). $dec\_p$ has outgoing transitions to all states that have same parent as $\gamma^\prime$ (added by line\# 16 of Algorithm~\ref{superHSM}). Thus the execution can be steered to $\gamma^\prime$ by providing the required input. (\RNum{2}) $parent(\sigma) \in parents(\sigma^\prime) \wedge parent(\sigma) \centernot = parent(\sigma^\prime)$. In this case, after the execution reaches the first $dec\_p$, it then can be moved using a series of $to\_child$ and $from\_parent$ transitions (line\# 20-26 of Algorithm~\ref{superHSM}) until reaching $\gamma^\prime$ whose current state is $\sigma^\prime$. 
(\RNum{3}) $parent(\sigma^\prime) \in parents(\sigma) \wedge parent(\sigma) \centernot = parent(\sigma^\prime)$. In this case, after the execution reaches the first $dec\_p$, it can then be moved using a series of $to\_parent$ and $from\_child$ transitions (lines\# 20-26 of Algorithm~\ref{superHSM})  until reaching $\gamma^\prime$ whose current state is $\sigma^\prime$.  Based on (\RNum{1}), (\RNum{2}), and (\RNum{3}), the proof of Proposition~\ref{prob:reach1} is complete.
\end{proof}

\end{prop}

\begin{prop}(Reachability of Transitions) 
\label{prob:reach-transition}
Let $\gamma$ be the current configuration of $L_r$, where $\gamma.\sigma \in \mathcal{S}_b$. By injecting $dbg$ and related messages, a sequence of  execution steps can be taken to execute the action of any transition, except for initial transitions and transitions starting from choice-points.

\begin{proof} (Reachability of transitions) Based on the execution semantics of \codee{HSM}s a prerequisite for the execution of the action of transition $t$ is to move the execution to a configuration $\gamma$ whose current state is (1) the source state of transition $t$, or (2) a basic state inside the composite state which is the source of transition $t$ (when a transition $t$ start from a composite state). According to Proposition~\ref{prob:reach1}, this can be done by injecting a $dbg$ message and providing proper inputs. Thus, we have to show that after reaching the source state of the transition $t$ according to (1) or (2), an execution step can be taken to execute the action of the transition which includes five cases based on the source state of transition $t$: (1) pseudo-state except for choice-points and initial states, (2) basic state,  (3) composite state, (4) choice-points, (5) initial state. \\
(\RNum{1}) The proof for Case-1 is trivial. According to the Rule-1 (see Def.~\ref{sec:semantic}), the execution step is taken from these states if there is an outgoing transition originating from them. Thus, an execution step can be taken that executes the action of transition $t$.\\
(\RNum{2}) The poof of case-2 and case-3 is similar to Case-1 assuming proper input messages are provided (trigger of transition $t$). As we discussed in Sec.~\ref{sec:debuggingprobe}, we provide a message injection feature that simplifies this. \\
(\RNum{3}) Case-4 is not part of the proposition, because it is not possible to ensure that the transition $t$ is executed due to its guard expression. Any buggy guard statement can prevent the execution of the transitions originating from a choice-point. \\
(\RNum{4}) Case-5 is not part of the proposition. There is no way for the execution to re-visit the transition starting from the initial state, except by restarting the execution. Based on (\RNum{1}) and (\RNum{2}), proof of Lemma~\ref{prob:reach-transition} is complete.

\end{proof}

\end{prop}

\subsubsection{Progress of the execution}

\begin{prop}{(Progress of the Execution)}
The execution of $\Re^{HSM}$ never reaches a stuck configuration assuming proper inputs are provided.

\begin{proof} (Sketch)
As we discussed in Sec.~\ref{sec:refinment},  the execution gets stuck due to two groups of issues: (1) Missing/problematic specification. (2) Missing input messages. All of the elements in group (1) are fixed by the refinement. Also, $dbg$ and other relevant messages can be injected by users which prevents a component from getting stuck because of missing inputs. Here, we do not present a detailed proof, but it can be performed similar to the previous proofs.

\end{proof}
\end{prop}

\subsection{Empirical Evaluation}

This section details experiments we conducted to assess the performance and overhead of our approach. In the following, we describe use-cases, evaluation metrics, experiments, and results.



\subsubsection{Use-cases}
To perform experiments, several use-cases are used. As shown in Table~\ref{table:usecases}, models have different complexities that range from simple models containing $11$ states to models with $350$ states. Simple models include the Car Door Central Lock system and the Digital Watch. The Car Door Central Lock system is a control system for locking and unlocking car doors. The Digital Watch is an implementation of classical digital watch, which is described in~\cite{harel1987statecharts}.

  \begin{table*}[t!]
    \caption{Model Complexity of Use-cases, Worst Case Transformation and Analysis Time}
    \centering
    \begin{tabular}{|c|c|c|c|c|c|c|c|c|c|}
    \hline
    \multirow{2}{*}{\textbf{Model}} & 
    \multicolumn{3}{|c|}{\textbf{Orig. Model Complexity}}  & \multicolumn{3}{|c|}{\textbf{Analysis Time (ms)}} &
    \multicolumn{3}{|c|}{\textbf{Transformation Time (ms)}} 

    \\ 
    \cline{2-10}
    & C & S & T & Median & Max. & Min. & Median & Max & Min
    \\
    \hline
    \hline
    Car Door Central Lock & 5 & 11 & 15   & 418  & 925 & 250      & 2024   & 3704   & 782 \\ \hline
    Digital Watch & 9 & 47 & 57           & 717  & 1535 & 322            & 5219   & 11126   & 2225\\ \hline
    Parcel Router & 8 & 14 & 25           & 418 & 1674 & 220           & 2877   & 5305   & 1279 \\ \hline
    Rover & 6 & 16 & 21                   & 604 & 925 & 313                    & 3062   & 5001   & 1254 \\ \hline
    FailOver  & 7 & 31 & 43               & 739 & 2247 & 257               & 4523   & 10416   & 1685 \\ \hline 
    Debuggable FailOver  & 8 & 350 & 620  & 1694 & 8454 & 347       & 13376  & 35500 & 1887\\ \hline
    \multicolumn{9}{l}{}\\
    \multicolumn{9}{l}{\footnotesize \textit{C}: Component, \textit{S}: State, \textit{T}: Transition, \textit{Orig.}: Original} 
    \end{tabular}
    \label{table:usecases}
  \end{table*}

The Parcel Router~\cite{1982-Swartout,1999-Magee} is an automatic system where tagged parcels are routed through successive chutes and switchers to a corresponding bin. The system is time-sensitive and jams can appear due to variations in the time required by a parcel to transit through the different chutes. It checks for potential parcel jams, and prevents parcels from being transferred from one chute to another until the next chute is empty. The simplified version ignores jams.

The Rover system model~\cite{ahmadi2016run,bagherzadeh2019model} allows an autonomous robot to move in different directions. It is equipped with three wheels, driven by two engines. It can move forward, move backward, and rotate. Additionally, it is equipped with several sensors, such as temperature and humidity sensors, to collect data from the environment, and an ultrasonic detection sensor, to detect and avoid obstacles.

The FailOver system~\cite{failover2,failoverUMLRT} is an implementation of the fail-over mechanism. It involves a set of servers processing client requests. To meet high availability, the system supports two replication modes, passive and active~\cite{failover}. In passive replication, one server component works as the master, handling all the client requests while backup servers are mainly idle, except for handshake operations. Whenever a malfunction occurs, resulting in a failure of the master server, a backup server is ranked up as the new master. In active replication, client requests are load-balanced between several servers. 

The Debuggable FailOver system is a debuggable version of the FailOver system, which is generated using MDebugger~\cite{mojtabadebugging}. The complexity of this model is high, and allows us to check that the refinement and analysis time do not skyrocket when the model size grows exponentially.

 \subsubsection{Evaluation Metrics}
We formulated the following metrics to assess the practicality of our approach.

\noindent
\textbf{Metric 1 (Performance of Analysis and Refinement).}  We use model analysis and transformation to fix partial models for the execution. The analysis and refinement are the core of our approach, and their performance is a crucial metric for the practicality of our approach. Thus, this metric measures the time required for the analysis and transformation of models. \\

\noindent
\textbf{Metric 2 (Overhead of refinement).} As discussed, the refinement adds certain elements to fix the execution of partial models. These new elements increase the complexity of the models in terms of the number of components, states, and transitions. This metric first measures how the complexity of refined models changes in comparison with the original ones. The refined model is created temporarily before execution, and is only used for code generation. Thus, this metric also measures the code generation time for original and refined models, in order to determine the side effects of the increased model size.

\noindent
\textbf{Metric 3 (Performance of the debugging probe).}  When executing the partial models, the execution of HSM is passed to the debugging probe, to read and apply user input. In the interactive model, there is always a delay imposed by users in the loop, and the performance is not an important factor. However, in the batch mode, it is essential that the debugging probe efficiently selects and applies the execution rules. This metric measures the time required to load, select, and parse rules.

\subsubsection{Experiments} In the following, we discuss the experiments used to calculate the metrics.\\

\noindent \textbf{Measuring the performance of static analysis and model transformation (EXP-1).} 
To effectively measure the performance of analysis and transformation, first we used Epsilon~\cite{epsilon2008} to create nine versions of each model (partial versions), listed in Table~\ref{table:usecases} by removing 10\%-90\% of their elements, randomly. This results in 60 models (including the original ones). In the rest of this section, we refer to these versions by merely mentioning the model name appended with the percentage of removed elements (e.g., Rover\%10 refers to a version of the model of the Rover system that has 10\% of its elements removed randomly). Also, we use the percentage without a model name to refer to all models with the same level of missing elements (e.g., 10\% refers to model versions of all use-cases that have 10\% of their elements missing).   

Second, we ran the model analysis and refinements 20 times against the original and their partial versions, with a configuration in which all components are assumed to be partial. The rationale for the configuration is to measure the performance in the worst-case scenario. As discussed, the refinement and analysis of a \codee{partial} components is much more expensive than the \codee{complete} and \codee{absent/ignored} components. No refinement is applied on \codee{complete} components, and the behaviour of an \codee{absent/ignored} component is replaced with a simple generic state machine. We recorded the time required for analysis and refinement, which is a reflection of their performance in the worst-case scenario. We also saved the partial and refined versions of the model that are used in \codee{EXP-2}.

\noindent \textbf{Measuring the overhead of the refinement (EXP-2)} 
First, we measure the complexity of the models, and their refined version resulting from \codee{EXP-1} in terms of the number of components, states, and transitions. Second, we generated code from them 20 times, and recorded the execution time of the code generation. This experiment reveals how the model complexity is increased when applying refinement, and what the effects of this increase are on the code generation.

\noindent \textbf{Measuring the performance of execution rule selection and application (EXP-2)}
To measure the loading/selection time of the execution rules, we generated 10,000 rules with 100 Lines of Code (LOC) in the context of the \codee{ABM} system which is a controller of an Automatic Banking Machine (ABM) designed using UML-RT. We performed a test that loads the rules in four scenarios, in which 10, 100, 1000, 10,000 rules are used accordingly. We recorded the loading time in each scenario. Then, using a test program, we called the rule selection method for the random context based on the \codee{ABM} system 1000 times, and measured the rule selection times.

To measure the time required to apply execution rules, we randomly generated four execution rule bodies, containing 1, 10, 100, 1000 lines in the context of the \codee{ABM} system. We ran a test to measure the time required to parse the rule bodies, 20 times. We did not measure the execution time of the rules' body, since their execution time is dependent on their content, which is controlled by users. The debugging probe executes the body of the rules as they are.  

\subsubsection{Setting and Reproducibility of Experiments}
We used a computer equipped with a 2.7 GHz Intel Core i5 and 8GB of memory, for all experiments, which is typical development PC. The experiments are automated using bash scripts. The scripts and models are publicly available at~\cite{pmdebuggersource} and can be used to repeat our experiments. Note that we intentionally used a standard computer comparable to those used by developers, rather than more powerful hardware, because the debugging of partial models typically needs to be carried out daily.

\subsection{Results}
\textbf{Metric 1 (Performance of analysis and refinement).}
Based on the result of \codee{EXP-1}, the \codee{Analysis Time} and \codee{Transformation Time} columns of Table~\ref{table:usecases} show the median, maximum and minimum time required to analyze and transform the ten versions of each use-case. For the largest model (Debuggable failover), the medians of analysis and transformation are less than two and 14 seconds, respectively. It is therefore safe to conclude that the performance of analysis and refinement is reasonable even when the configuration of all components is set to \codee{partial} which is the worst-case configuration. Typically, the execution of partial models is focused on executing specific components, and the rest of the components are assumed to be \codee{complete} or \codee{absent/ignored} which is less expensive to analyze and refine. 

\begin{figure*}
    \centering
    \includegraphics[width=8cm]{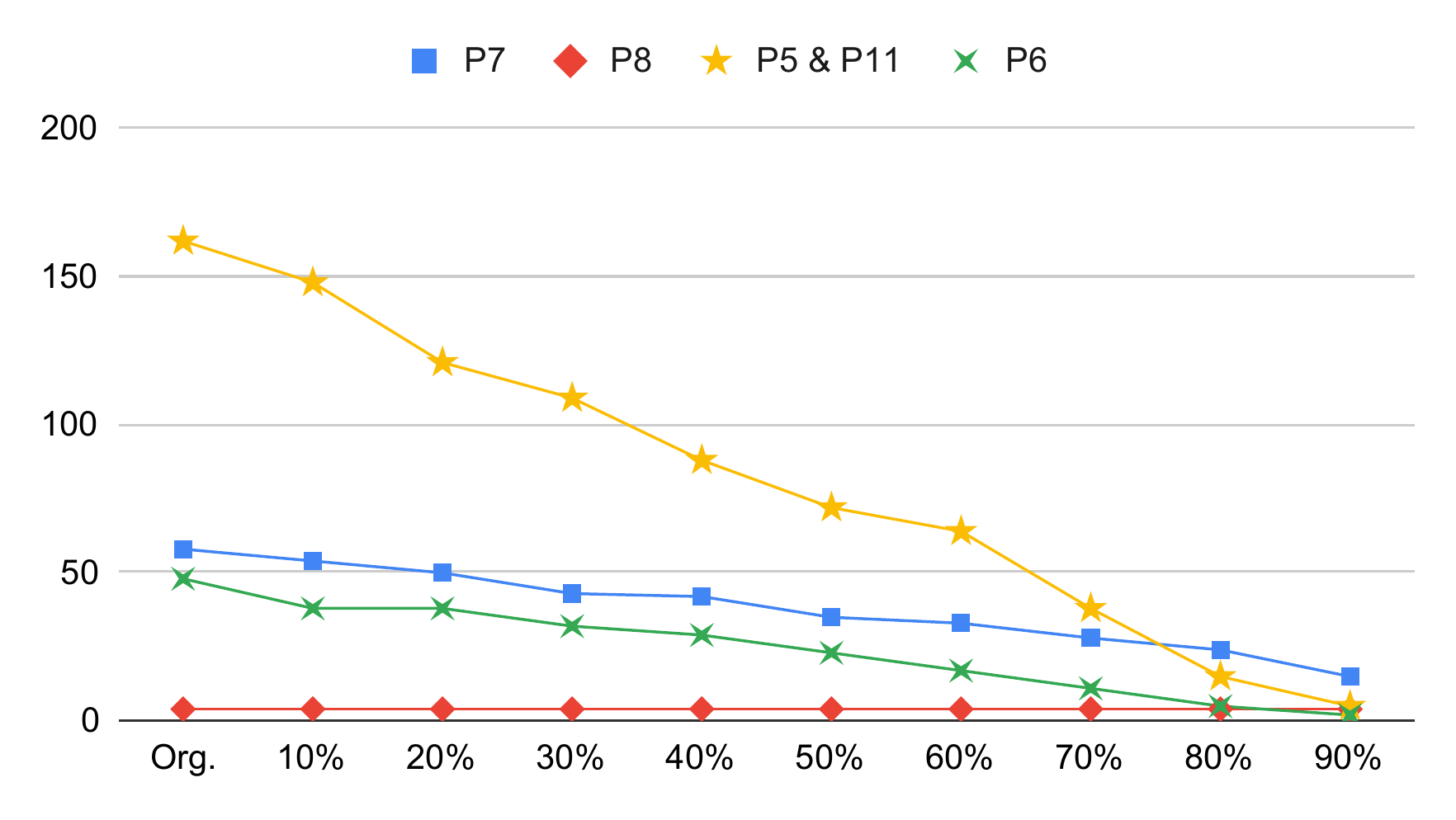}
    \includegraphics[width=8cm]{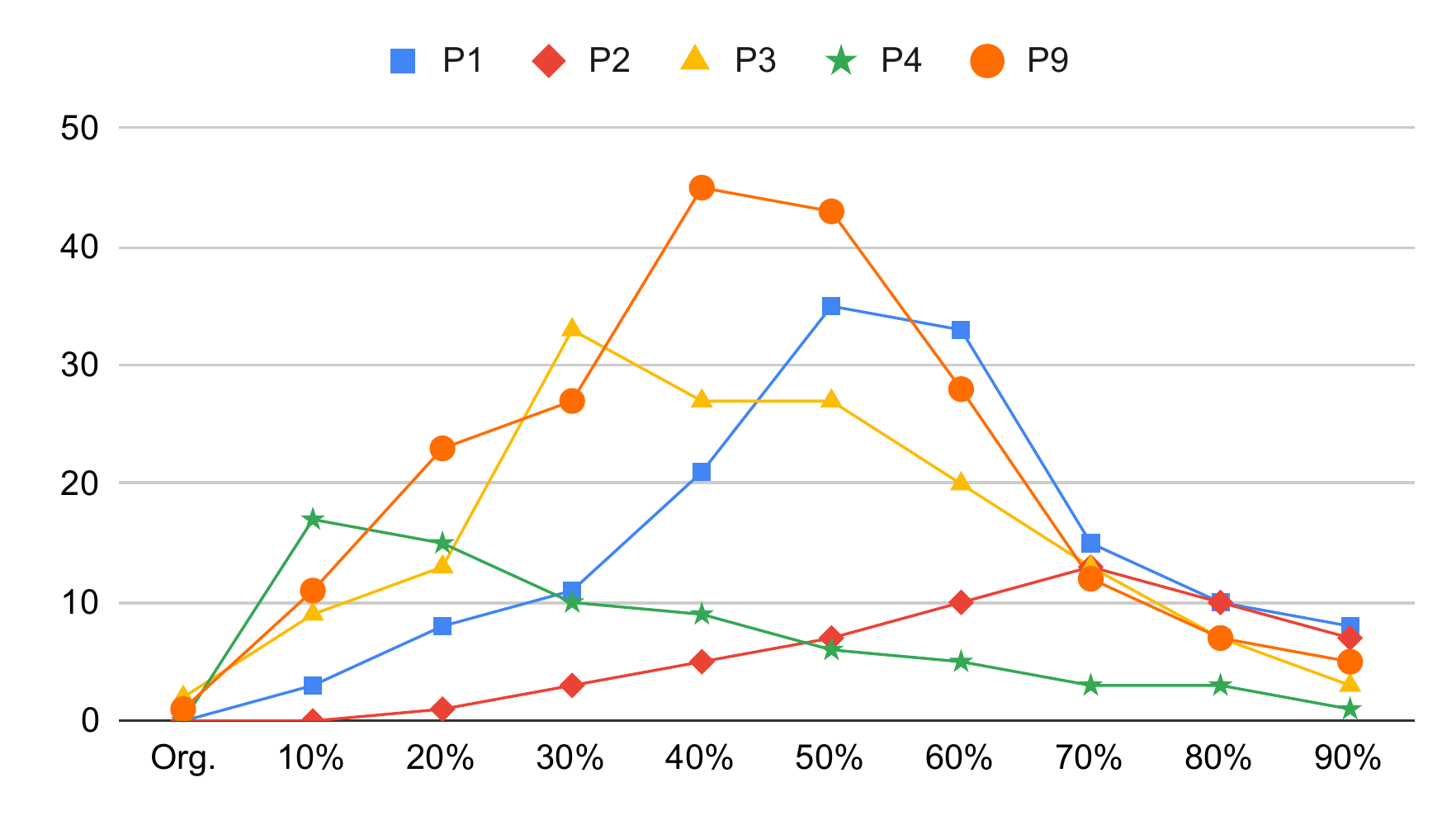}
    \caption{Number of elements (\textit{y} axis) in sets P1-P11 for versions (Orig. and 10\%-90\%) of Debuggable FailOver system}
    \label{fig:problematicelements}
\end{figure*}

\begin{figure*}[!ht]
\centering
\begin{subfigure}[b][7cm][b]{.48\textwidth}
\centering
\includegraphics[width=8cm]{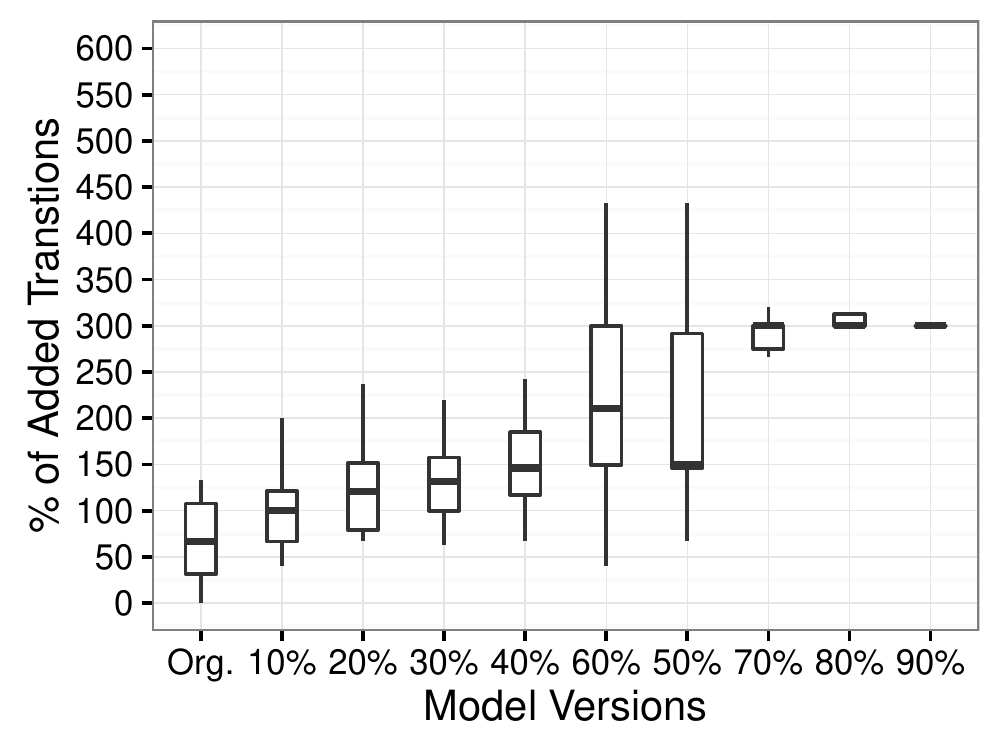}
\end{subfigure}
\begin{subfigure}[b][7cm][b]{.48\textwidth}
\centering
\includegraphics[width=8cm]{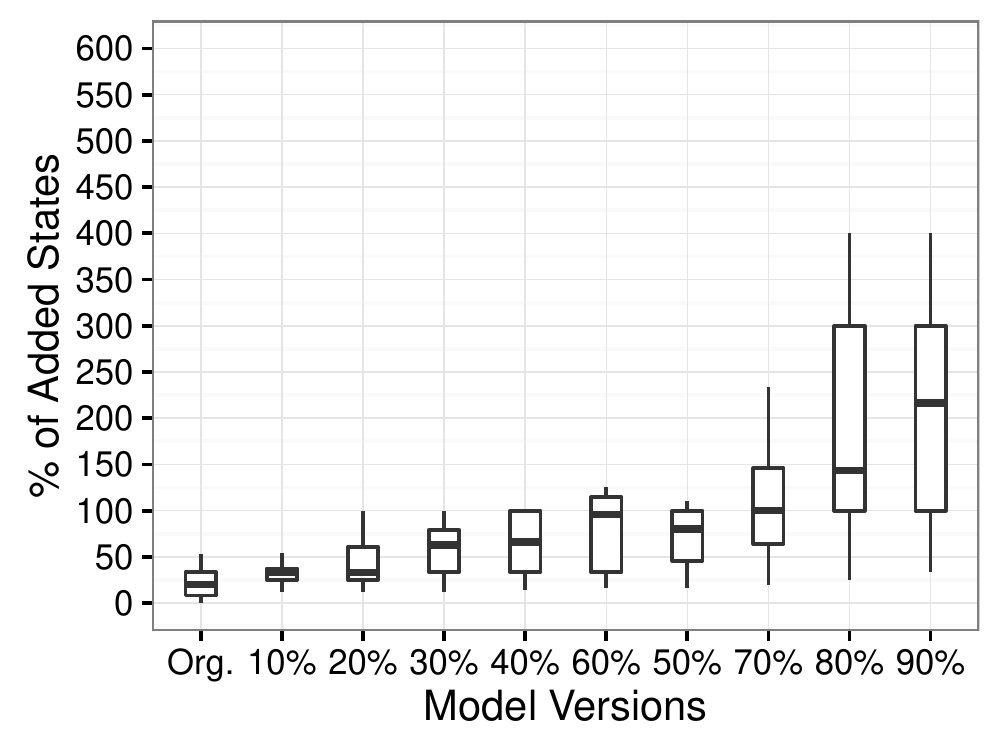}
\end{subfigure}%
\caption{Increased complexity of refined models (original and partial versions)}
\label{fig:complexityoverhead}
\end{figure*}

Figure~\ref{fig:problematicelements} shows number of elements in P1-P11 except P10 for the different versions (Orig.~and 10\%-90\%) of the Debuggable FailOver system. The number of elements in \textit{P5}, \textit{P6}, \textit{P7}, \textit{P8}, \textit{P11} decreases as the number of removed elements increases, i.e., the number of elements in these sets are highest for original models. This is because of the overestimation for extracting these sets which is based on the numbers of elements in the HSM. For example, \textit{P11} includes all basic states of the \textit{HSM}. Thus removing more elements (states and transitions) in the \textit{HSM} causes a decrease in the number of elements in \textit{P11}. However, the number of elements for \textit{P1},  \textit{P2}, \textit{P3}, \textit{P4}, \textit{P9} reaches its maximum  between versions 30\%- 60\%, because in these cases the removed elements cause a maximal amount of issues for the remaining elements of the \textit{HSM}. This number then decreases in the subsequent versions when more and more of these remaining elements are also removed. Note that the number of elements in \textit{P10} is not included in the figure, because we only remove states and transitions from the original model. Thus, the number of elements in \textit{P10} does not change between versions.

\noindent\textbf{Metric 2 (Overhead of the refinement).} Based on the results of \codee{EXP-2}, Fig.~\ref{fig:complexityoverhead} shows the percentage of added elements (states and transitions) to the original models and their partial versions, during the refinement (i.e., the number of the added element divided by the number of elements before refinement multiplied by 100). 
Not surprisingly, the number of added elements increases as the number of removed elements from models increases. Removing more elements introduces more problems for the execution, which in turn requires more elements to be added to fix these problems. The percentage of added states is between 20\% (the median of the percentage of added states for original versions of models) and 216\% (the median of the percentage of added states for model versions with 90\% removed elements). The percentage of added transitions is between 67\% (median of the percentage of added transitions for the original model versions) and 300\% (median of the percentage of added transitions for model versions with 90\% removed elements). Note that the percentage of added transitions for the versions with 90\% removed elements is almost fixed (300\%) because almost all are removed and the refinement always adds almost the exact same elements to refine them similar to \codee{absent/ignored} components. The percentage of added transitions is higher than the percentage of added states, since many of the execution problems are fixed by adding transitions. Also, the number of components increases only by one (i.e., the \codee{dbg\_agent} component). We argue that these overheads are reasonable compared to the capabilities provided by the refined models, for the following reasons:
\begin{itemize}
    \item In most of the cases, the refinement adds elements when there is a missing/problematic element and there is no other way to fix them using existing tools and techniques. Our approach simply automates the fix for problematic elements. Otherwise, users have to fix them manually, which is time-consuming and tedious. 
    \item The refined models are temporary models, which are only used for code generation. Thus, the overhead of added elements has no side effect except for the code generation. The result of the second part of \codee{EXP-2} shows that the code generation of the refined model is only 8\% slower than the code generation for original models, which is calculated based on the median of the time for code generation from refined models, divided by the time for code generation from the original models.  
    \item As discussed, the experiments are performed using the worst-case configuration, and their results reflect the maximum costs of our approach. Otherwise, using realistic configurations, which focus on the execution of certain components, can even decrease the complexity of the refined models with respect to the original models. E.g., the refinement of the Debuggable Failover system by setting the completion level of the \codee{Client} component to \codee{partial} and the level of the other  components to \codee{absent/ignored} results in a refined model with 138 states and 326 transitions, which is almost 50\% smaller than the original model!
\end{itemize}

\begin{table}[!t]
 \begin{threeparttable}
\centering
	\caption{Required time for loading and selection of execution rules} 
	\label{tab:debuggingprobe}
    \begin{tabular}{p{1cm}p{1.3cm}p{2cm}p{2cm}}
    \hline
     \small \textbf{\# of Rules} & \textbf{LOC of Rules} & \small \textbf{Loading time (ms)}  & \textbf{Selection time (ms)} \\ 
    \hline
    10&    100 &  3    &     0.006 \\
    100&   100 &  13   &     0.006 \\
    1000&  100 &  96   &     0.006 \\
    10,000& 100 &  950  &     0.010 \\
   \hline
    \end{tabular}
\end{threeparttable}
    \bigskip 
\end{table}

\noindent\textbf{Metric 3 (Performance of the debugging probe).}
Based on the result of \codee{EXP-3}, Table~\ref{tab:debuggingprobe} shows the time required for loading and selecting rules by the debugging probe. The selection time is the median time of rule selection for 1,000 times. The loading of rules occurs only once the execution of the system starts. During the loading, the script of execution rules is loaded and parsed. The parsing in this phase only parses the rules’ header, and saves their body as text. As shown in Table~\ref{tab:debuggingprobe}, the debugging probe can load 10,000 rules in less than a second, which is acceptable, because it happens only once. 

As discussed in Sec.~\ref{sec:debuggingprobe}, in batch mode execution, the debugging probe must select an applicable rule from the defined rules whenever the execution reaches the decision points. As shown in Table~\ref{tab:debuggingprobe}, the rule selection time is negligible (less than a millisecond), and it is, therefore, safe to conclude that rule selection performance is acceptable. 

When an execution rule is selected, the debugging probe parses the rule's body and executes it. Thanks to ANTLR~\cite{antlr}, the parsing time of the rule's body has reasonable performance. Rule bodies with 1-1000 LOC can be executed in less than a second (the median execution time for a rule body with 1000 LOC is 550 milliseconds).

According to the results mentioned above (i.e., acceptable performance of analysis, refinement, and debugging probe and reasonable overhead of the refinement), we conclude that our approach is a practical approach for the execution and debugging of partial models.

\section{Discussion}
\label{sec:discussion}
In the following we discuss issues with the input-driven execution of partial models, alternative solutions for the refinements of partial models, and threats to the validity of this work.

\subsection {Issues with the input-driven execution}

Since the execution rules are defined using a scripting language, it may contain bugs similar to any other scripting language. Generally, there is no solution to this problem, and to mitigate this issue partially, our refinement method guarantees behavioral preservation (see Sec.~\ref{sec:proofs}). Therefore, it cannot introduce new bugs into the completed part of the models. Also, to help users when the script is buggy, the execution engine switches back to interactive mode and allows users to provide a correct input.  Finally, providing proper tooling such as a high-quality editor for writing and validating scripts can mitigate the challenge of correct script authoring.

\subsection{Alternative refinement solutions}
Note that the proposed refinement (see Sec.~\ref{sec:refinment}) approach is devised experimentally by experimenting with and evaluating possibly many different solutions. In Section~\ref{sec:evaluation}, we discussed the correctness of our refinement approach concerning the relevant constraints (see Sec.~\ref{sec:framework}). We also showed that our approach has reasonable performance and overhead for the refinement of partial models. However, at this stage, we do not claim that our refinement approach is the optimal solution, and alternative and even better refinement approaches can be proposed, especially if certain trade-offs or assumptions are made.  For example, the refinement can only focus on fixing specific problems rather than addressing all of them depending on the users' needs, e.g., refining only elements that participate in the lack of progress can be simpler and faster than our current approach. Nevertheless, the proposed refinement is comprehensive and can be used as a reference method to devise more specialized methods targeting more specific problems. 


\subsection{Threats to the Validity}

\textbf{Internal threats.} (1) To evaluate this work, we use generated models rather than using real partial models. Thus, the results of our evaluation may not be generalized to real partial models. To mitigate this issue, we tried to generate models with a wide range of partialness over several case studies to make sure they are representative of typical partial models. Also, we used a worst-case configuration for the evaluation. (2) The implementation of our approach is not trivial, and therefore our implementation may have bugs. To mitigate this problem, we rely on well-known tools and frameworks, such as EMF and Epsilon. We also have performed a thorough test and validation of our implementation. (3) The online survey participants viewed only a short demonstration video to familiarise themselves with the execution of partial models and our proposed approach. We partially mitigate this issue by carefully designing the video and targeting the MDE experts as participants, most of them already aware of the problems surrounding the execution of partial models. (4) Our study may be designed in a way that, inadvertently, steers participants towards specific answers. To mitigate this issue, the participants are allowed to provide other answers rather than what is presented to them.  Also, providing an email address is not mandatory to enable participants to be anonymous in case they want to provide negative feedback.

\textbf{External threats.} We targeted MDE experts as our survey participants to make sure they can provide us relevant and high-quality feedback. However, this may also be a threat to the survey's results since some of the participants may be biased about how, e.g., the problem of executing partial models should be dealt with.

\section{Related Work}
\label{related-work}
A large amount of related work exists, and only the most relevant can be discussed here. Existing work can be divided into three categories: (1) work on model-level debugging and execution, (2) work on partial models, which tries to address specification, analysis, and transformation of partial models, and (3) work on partial programs that deals with the  parsing, analysis, and completion of partial programs in the context of different programming languages. 

\noindent\textbf{Model-level execution and debugging.}
Existing techniques of model execution are based on either interpretation or translation. Interested readers can refer to~\cite{Ciccozzi2018} which provides a comprehensive survey of existing work in the context of the model execution.

Model-level debugging techniques can be classified into interactive debugging and debugging by tracing. Interactive debugging allows users to directly investigate and modify the execution of models during the execution, by providing features such as setting breakpoints and stepping over the execution. Interactive model-level debugging is supported by several MDD tools, e.g., Matlab StateFlow~\cite{Stateflow}, AF3~\cite{AutoFocus}, xtUML~\cite{xtUML} and YAKINDU~\cite{YAKINDU}. We also presented a new approach for supporting interactive model-level debugging in~\cite{MDebuggerDemo,mojtabadebugging} by using model transformation techniques. 

In debugging by tracing, the model or the generated code is instrumented to generate useful execution traces. Then, the traces are collected and used for offline analysis and debugging. Hojaji et al.~\cite{Hojaji2019} surveys the existing work in the context of model execution tracing. Examples of existing work and MDD tools supporting trace analyses via code instrumentation include~\cite{haberl2010model,thane2003replay,hili2020model,Iyenghar:2010:MBA:1879021.1879031,iyenghar2011model,gery2002rhapsody}. For instance, Iyengar et al.~\cite{Iyenghar:2010:MBA:1879021.1879031,iyenghar2011model,graf2006dynamic} propose an optimized model-based debugging technique for RTE systems with limited memory. They use a monitor on the target platform to collect the generated traces and a debugger (executed on a host with sufficient memory) to analyze the traces offline, and to display results on the model elements. Das et al.~\cite{das2016supporting} propose a configurable tracing tool based on LTTng. They rely on code instrumentation to produce tracepoints useful for LTTng. 

To the best of our knowledge, none of the existing work in the context of model execution and debugging supports the execution and debugging of partial models.
    
\subsection{Partial models}
In the context of MDD, the partial models are mainly used to deal with uncertainties of type `known unknown'. Existing research proposes mechanisms to define partial models using relaxed meta-models~\cite{sen2012using}, model annotation~\cite{famelis2012partial}, UML profiles~\cite{zhang2016uncertainty}, and graphical notations~\cite{famelis2013mav}. They leverage the partial models for analysis~\cite{famelis2012partial,semerath2017graph}, requirement management and analysis~\cite{salay2013managing}, testing~~\cite{sen2012using,zhang2017uncertainty}, and bi-directional transformation~\cite{eramo2015managing}. Also, some research addresses the refinement~\cite{salay2012language,salay2015methodology},  transformation~\cite{famelis2013transformation} and completion~\cite{sen2007partial} of partial models. E.g., the wok~\cite{famelis2012partial,Famelis2017} presents a rich formalism for partial models, which marks model elements with four special annotations (\codee{may}, \codee{set}, \codee{variable}, and \codee{open}) with well defined semantics. They show how the partial models can be concretized into possible design candidates. Sen et al.~\cite{sen2012using} present 
a semi-automated tool that supports the specification and completion of partial models, which are used for the testing of model transformations. They show that the testing of model transformations using partial models is as effective as using human-made models.

To the best of our knowledge, no work in the context of partial models addresses the execution and debugging of partial models. Our work does not require specification of partial elements explicitly by users, since it detects all of them automatically by static analysis. Automatic detection of partial elements allows users to execute the models with minimum effort. Note that the partiality that our approach detects only concerns the execution, and may not be suitable for managing uncertainties in requirements or design models.

\subsection{Partial programs}
An extensive body of work exists for dealing with and leveraging partial programs. The most important of them can be classified as follows.

\noindent\textbf{(1) Parsing of partial programs} Typically existing compilers can handle only complete programs. As a partial program is a subset of a complete program, many of its variables' types and library calls are unknown. Thus, parsing partial programs requires extra effort, mainly for the inference of missing types, and resolving unknown function calls. E.g., Zhong et al.~\cite{Zhong:2017:BCT:3155562.3155646} propose an approach that resolves unknown types and function calls for partial Java programs by analyzing the existing complete program versions. Melo et al.~\cite{melo2017inference} present a technique to support the compilation of incomplete C code. 

Koppler~\cite{koppler1997systematic} presents a systematic approach to implement fuzzy parsers, which extract high-level structures out of incomplete or syntactically incorrect programs. Moonen~\cite{moonen2001generating} proposes a solution in the form of island grammars that partitions code into islands (recognizable constructs of interest) and water (remaining parts). Dagenais et al.~\cite{dagenais2008enabling} propose a framework that uses heuristics to recover the declared type of expressions and resolve ambiguities in partial Java programs. Note that since the models are saved in the form of an abstract syntax tree (AST), the need for this type of research is unnecessary in the context of MDD. 

\noindent\textbf{(2) Partial program analysis/verification to deal with poor scalability and missing components}  E.g., modular model checking, introduced in~\cite{Grumberg:1994:MCM:177492.177725}, verifies properties of system modules, under some assumptions about the environment. Colby et al.~\cite{Colby:1998:ACO:277652.277754} present an approach for automatically closing an open concurrent reactive system (i.e., a system with missing components) by generating an environment that can provide any input at any time to the system. This result is a self-executable system, which can exhibit all the possible reactive behaviors of the original system and therefore can be used for the state space exploration that is required for verification and analysis purposes. Our refinement of \codee{absent/ignored} components is similar to this work. 

\noindent \textbf{(3) Program synthesis techniques based on partial programs (synthesis by sketching).}
Instead of synthesizing a program from scratch, work in this category uses a partial program (i.e., a program with holes) along with a specification, test harness, or reference implementation, and tries to fill the holes using synthesis techniques. E.g., Solar-Lezama et al.~\cite{solar2005programming} introduce the concept of programming with sketches and
presents \codee{Stream Bit} as a new programming approach based on sketching. Existing sketching techniques (e.g., \cite{Solar-Lezama2013}) translate the partial program into a propositional satisfiability problem, and leverage counter-example-guided inductive synthesis to generate a program using existing SAT solvers. Hua et al.~\cite{hua2017edsketch} introduce \codee{EdSketch} that performs execution-driven sketching for synthesizing Java programs using a backtracking depth-first search. 

\section{Conclusion and Future Work}
\label{sec:conclusion}
In this paper, we have proposed a conceptual framework for the execution and debugging of partial models, which consists of \codee{static analysis}, \codee{automatic refinement}, and \codee{input-driven execution}. Using static analysis, we extract the problematic elements that prevent execution. The problematic elements are automatically fixed by adding decision points and related specifications into the partial models. Finally, the refined models are executed with the help of user input, either interactively or via a script. We have created a debugger for the debugging of partial UML-RT models (\codee{PMExec}) based on the proposed framework. We have applied \codee{PMExec} to the debugging of several use-cases, and have evaluated its performance for analysis, refinement, and handling of users input. Despite being a prototype, the performance of \codee{PMExec} is acceptable, which shows that our approach is a viable approach for the debugging of partial models.

We have made the implementation of \codee{PMExec} publicly available. The modeling community can extend it and use it for more research on, e.g., (1) the exhaustive execution of partial models for testing or run-time verification, (2) the synthesis of models by sketching, (3) using the proposed framework to support partial execution and debugging of partial models expressed in other modeling languages, and (4) automatic completion of missing specifications, rather than taking inputs from users.

\section*{Acknowledgment}
This work has been supported by Zeligsoft Ltd, Malina Software Corp., Cmind  Inc., and the Natural Sciences and Engineering Research Council of Canada (NSERC).
\ifCLASSOPTIONcaptionsoff
  \newpage
\fi
\bibliographystyle{IEEEtran}
\bibliography{sigproc.bib,IEEEabrv.bib}

\end{document}